# Exact Non-Identity Check and Gate-Teleportation-Based Indistinguishability Obfuscation Are NP-Hard for Low-T-Depth Quantum Circuits

Joshua Nevin

***Abstract*—In 2021, Broadbent and Kazmi developed a gate-teleportation-based protocol for computational indistinguishability obfuscation of quantum circuits. This protocol is efficient for Clifford+T circuits with logarithmically many T-gates, where the limiting factor in the efficiency of the protocol is the difficulty, on input a quantum circuit $C$, of the classical task of producing a description of the unitary obtained by conjugating a Pauli $P$ (corresponding to a Bell-measurement outcome) by $C$, where this description only depends on the input-output functionality of $CPC^\dagger$. The task above, in turn, is at least as hard as the problem of determining whether two $n$-qubit quantum circuits are perfectly equivalent up to global phase. In 2009, Tanaka defined the corresponding decision problem Exact Non-Identity Check (ENIC) and showed that ENIC is NQP-complete and thus NP-hard as well. Motivated by the 2021 protocol, we consider in this work what happens when we pass from low T-*count* to low T-*depth*. We show that, for Clifford+T-circuits of T-depth $O(\log(n))$, deciding ENIC remains NP-hard. In particular, we show this by relating certain decision problems on Pauli coefficients of Clifford+T unitaries to well-known hardness results for codeword weights in binary linear codes. This effectively rules out the possibility, for Clifford+T-circuits of logarithmic T-depth, of either efficient ENIC or efficient gate-teleportation based computational indistinguishability obfuscation, unless P = NP. It also provides a new, independent proof of the hardness of ENIC, even at low T-depth. Along the way, we prove the NP-hardness of some related circuit problems.**

***Index Terms*—Quantum circuits, cryptography, obfuscation, binary codes, NP-hardness.**

## I. Preliminaries on Quantum Gates and Circuits

To FORMALLY introduce our main results in Section IV, we first review some preliminaries on quantum circuits in Section I and we review some preliminaries on both classical and quantum complexity and cryptography in Section II. Then, in Section III, we review the gate-teleportation-based indistinguishability obfuscation protocol of [1]. We first review the following quantum notions.

Received 15 December 2025; revised 18 June 2026; accepted 9 August 2026. Date of publication 13 August 2026; date of current version 17 September 2026. This work was supported in part by the Digital Horizon Europe Project Foundations of Quantum Computational Advantage (FoQaCiA) under Grant 101070558 and in part by the Natural Sciences and Engineering Research Council of Canada (NSERC) under Grant ALLRP-578455-2022.

The author is with the Department of Mathematics and Statistics, University of Ottawa, Ottawa, ON K1N 6N5, Canada (e-mail: janevin@uwaterloo.ca).

*Definition 1:*

1) $(\mathbb{C}^2)^{\otimes n}$ denotes the $2^n$-dimensional vector space over $\mathbb{C}$ spanned by computational basis states $\{|x\rangle : x \in \{0,1\}^n\}$. In general, for $|\phi\rangle \in (\mathbb{C}^2)^{\otimes n}$ and $|\psi\rangle \in (\mathbb{C}^2)^{\otimes m}$, we write $|\phi\rangle|\psi\rangle$ to mean the tensor product $|\phi\rangle \otimes |\psi\rangle \in (\mathbb{C}^2)^{\otimes(n+m)}$.
2) The *trace distance* between two $n$-qubit density matrices $\rho, \sigma$ is given by
$$\|\rho - \sigma\|_{\text{tr}} := \frac{1}{2}\text{tr}\left(\left| \sqrt{(\rho-\sigma)^\dagger(\rho-\sigma)}\right|\right)$$
where $|\sqrt{A}|$ denotes the positive square root of positive semi-definite matrix $A$.
3) An operator $\Phi : (\mathbb{C}^2)^{\otimes n} \to (\mathbb{C}^2)^{\otimes m}$ is called *type* $(n,m)$. We say that $\Phi$ is *admissible* if its action on density matrices is linear, trace-preserving, and completely positive.
4) The *diamond norm* between admissable operators $\Phi, \Psi$, both of type $(n,m)$, is given by
$$\|\Phi - \Psi\|_\diamond := \max_{\rho \in (\mathbb{C}^2)^{\otimes 2n}} \|(\Phi \otimes \mathsf{I}^{\otimes n})\rho - (\Psi \otimes \mathsf{I}^{\otimes n})\rho\|_{\text{tr}}$$
where the maximum is taken over all $2n$-qubit density matrices.

A quantum circuit is a directed acyclic network of unitary gates connected by wires, where the wires represent the path of the qubits over time. In general, a quantum circuit can map $n$ qubits to $m$ qubits (e.g. by having ancilla inputs). Analogous to Definition 1, we say that a quantum circuit is of *type* $(n,m)$ if it maps $n$ qubits to $m$ qubits. Two quantum circuits $F, F'$ of the same type are called *perfectly equivalent* if $\|F - F'\|_\diamond = 0$. A quantum circuit is called *reversible* if it is of type $(n,n)$, i.e. it implements a unitary transformation on $(\mathbb{C}^2)^{\otimes n}$. We call such a circuit an *n-qubit* (quantum) circuit. We now recall the following 1- and 2-qubit unitary gates.

*Definition 2:*

$$\mathsf{X} := \begin{pmatrix} 0 & 1 \\ 1 & 0 \end{pmatrix} \qquad \mathsf{Z} := \begin{pmatrix} 1 & 0 \\ 0 & -1 \end{pmatrix}$$
$$\mathsf{Y} := i\mathsf{XZ} = \begin{pmatrix} 0 & -i \\ i & 0 \end{pmatrix} \quad \mathsf{H} := \tfrac{1}{\sqrt{2}}\begin{pmatrix} 1 & 1 \\ 1 & -1 \end{pmatrix}$$
$$\mathsf{S} = \sqrt{\mathsf{Z}} := \begin{pmatrix} 1 & 0 \\ 0 & i \end{pmatrix} \quad \mathsf{T} = \sqrt{\mathsf{S}} := \begin{pmatrix} 1 & 0 \\ 0 & e^{i\pi/4} \end{pmatrix}$$

$$\mathsf{CZ} := \begin{pmatrix} \mathsf{I} & \mathbf{0} \\ \mathbf{0} & \mathsf{Z} \end{pmatrix} = \begin{pmatrix} 1 & 0 & 0 & 0 \\ 0 & 1 & 0 & 0 \\ 0 & 0 & 1 & 0 \\ 0 & 0 & 0 & -1 \end{pmatrix}$$

Next, we recall the following groups and introduce some additional notation for circuit classes and gates.

*Definition 3:*

1) We let $\mathcal{P}_1$ denote the 1-qubit Pauli group $\bigcup_{1\le k\le 4}\{i^k\mathsf{I}, i^k\mathsf{X}, i^k\mathsf{Z}, i^k\mathsf{Y}\} = \langle \mathsf{X},\mathsf{Y},\mathsf{Z}\rangle$ and we let $\mathcal{P}_n$ denote the $n$-qubit Pauli group, i.e. the $n$-fold tensor product of $\mathcal{P}_1$. We let $\mathcal{U}(2^n)$ denote the group of $2^n\times 2^n$-unitaries in $M_{2^n\times 2^n}(\mathbb{C})$. We refer to the elements of $\mathcal{P}_n$ as $n$-qubit *Paulis* (or just Paulis if $n$ is clear).
2) We recall that the $n$-qubit *Clifford group* is $\mathcal{C}_n := \{C \in \mathcal{U}(2^n) : C\mathcal{P}_nC^\dagger \subseteq \mathcal{P}_n\}$, i.e. the normalizer of $\mathcal{P}_n$. The phaseless Clifford group $\mathcal{C}_n/\langle e^{i\theta}\mathsf{I}^{\otimes n} : \theta\in\mathbb{R}\rangle$ is finite and, in particular, generated by tensor products of $\mathsf{H},\mathsf{S},\mathsf{CZ}$ acting on individual qubits (resp. pairs of qubits). By a *Clifford circuit* we mean a reversible quantum circuit with gates drawn from $\{\mathsf{X},\mathsf{Z},\mathsf{H},\mathsf{CZ},\mathsf{S}\}$.
3) By a Clifford+T circuit we mean an $n$-qubit circuit $C$ with gates drawn from $\{\mathsf{X},\mathsf{Z},\mathsf{H},\mathsf{CZ},\mathsf{S},\mathsf{T}\}$. The T-*count* of $C$ is the number of T-gates that appear in $C$, and the T-*depth* of $C$ is the number of time steps in which at least one T-gate appears.
4) We let $\mathcal{T}_n$ denote the set of $n$-qubit Clifford+T-circuits. For $d\ge 0$, we let $\mathcal{T}_n^d$ denote the set of elements of $\mathcal{T}_n$ of T-depth *at most* $d$ and we let $\mathcal{T}^d := \bigcup_{n\ge1}\mathcal{T}_n^d$. In particular, the set of $n$-qubit Clifford circuits is $\mathcal{T}_n^0$. More generally, given a function $f:\mathbb{N}\to\mathbb{Z}_{\ge0}$, we write $\mathcal{T}^{f(n)}$ to denote the class of circuits $\bigcup_{n\ge1}\mathcal{T}_n^{f(n)}$.
5) Given a prespecified $n$ and $1\le j\le n$, and a single qubit gate $\mathsf{A}$, we let $\mathsf{A}_j$ denote $\mathsf{I}^{\otimes(j-1)}\otimes\mathsf{A}\otimes\mathsf{I}^{\otimes(n-j)}$. Likewise, for $1\le i<j\le n$, we let $\mathsf{CZ}_{i,j}$ denote the $n$-qubit circuit consisting of $\mathsf{CZ}$ acting on wires $i,j$ and identity on the remaining wires (recall that $\mathsf{CZ}$ is symmetric w.r.t the target and control qubits).

A set $S$ of quantum gates is called *universal* if, for any $n\ge1$, any $U\in\mathcal{U}(2^n)$ can be approximated to within arbitrary accuracy by an $n$-qubit circuit with gates drawn from $S$. It is a well-known fact that the Clifford gate set is not universal but that adding any non-Clifford gate, such as the T-gate, to the Clifford gate set results in a universal gate set, and the Clifford+T gate set is the most well-studied universal gate set. We now recall the standard convention for representing $n$-qubit Paulis by elements of $\mathbb{F}_2^{2n}$.

*Definition 4:* Given a vector $v=(a_1,\dots,a_n,b_1,\dots,b_n)\in\mathbb{F}_2^{2n}$, we set

$$\mathsf{P}^v := \left(i^{a_1b_1}\mathsf{X}^{b_1}\mathsf{Z}^{a_1}\right)\otimes\left(i^{a_2b_2}\mathsf{X}^{b_2}\mathsf{Z}^{a_2}\right)\otimes\cdots \otimes\left(i^{a_nb_n}\mathsf{X}^{b_n}\mathsf{Z}^{a_n}\right)$$

Thus, $\mathcal{P}_n=\{i^k\mathsf{P}^x : k\in\mathbb{Z}/4\mathbb{Z}$ and $x\in\mathbb{F}_2^{2n}\}$, and the phaseless group $\mathcal{P}_n/\langle\pm\mathsf{I}^{\otimes n},\pm i\mathsf{I}^{\otimes n}\rangle$ is isomorphic to $\mathbb{F}_2^{2n}$. The elements of $\{\mathsf{P}^x : x\in\mathbb{F}_2^{2n}\}$ form an orthonormal basis for $M_{2^n\times2^n}(\mathbb{C})$ with respect to the Hilbert-Schmidt inner product, so, given a $U\in M_{2^n\times2^n}(\mathbb{C})$, there is a unique expression of $U$ as a $\mathbb{C}$-linear combination of elements of $\{\mathsf{P}^x : x\in\mathbb{F}_2^{2n}\}$, which we call the *Pauli expansion* of $U$, where the coefficient of $\mathsf{P}^z$ is $\langle U,\mathsf{P}^z\rangle$. Studying the Pauli expansion of unitaries is a natural and fruitful way to analyze the complexity of the corresponding quantum operation (see, for example, [2] and [3]). Our interest in this paper lies in studying this expansion for low-T-depth quantum circuits, and relating it to some cryptographic tasks. More specifically, we study this expansion for the class of low T-depth circuits that arise from conjugating a Pauli by a Clifford+T circuit, which is considerably easier to study than general circuits. Cryptographically, this is a very natural class of circuits, corresponding precisely to the correction that arises when performing a desired computation on a quantum state which has been encrypted by a random Pauli (see Section III). As we refer to this class of circuits very often, we say that an $n$-qubit circuit is a *PC*-circuit if it is of the form $C\mathsf{P}^xC^\dagger$ for some $n$-qubit Clifford+T circuit $C$ and $x\in\mathbb{F}_2^{2n}$.

## II. Preliminaries on Algorithms and Obfuscation

Informally, *obfuscation* is the cryptographic task of rendering a computer program "unintelligible" in a certain sense while preserving its functionality. That is, given a circuit $C$, the goal is to output another circuit $C'$ such that $C'$ has the same input-output functionality as $C$ but, to the greatest extent possible, everything else about $C$ is hidden. An *obfuscator* is an algorithm $\mathcal{O}$ which performs this task. If the output circuit $C'$ does not reveal anything about $C$ except for its input-output functionality, then $\mathcal{O}$ is called a *virtual black-box obfuscator*. The concept of virtual black-box obfuscation was formalized in [4], and the same work also showed that virtual black-box obfuscation is not achievable in general. Motivated by this, [4] also introduced a weaker notion of obfuscation called *indistinguishability obfuscation*. Informally, an indistinguishality obfuscator is an algorithm $i\mathcal{O}$ which takes circuits as input and has the property that, if two circuits have the same input-output functionality, then their obfuscations are indistinguishable (see Section VII of [4]). As noted in [4], *inefficient* indistinguishability obfuscation is indeed possible, because, given a circuit $C$, we can output the unique lexicographically minimal circuit $C'$ among the circuits of size $|C|$ which compute the same function as $C$. The extent to which *efficient* indistinguishability obfuscation is possible has attracted a lot of attention because, as a cryptographic primitive, indistinguishability obfuscation can be used, among other things, for digital signatures, public key encryption (see [5]), and fully homomorphic encryption (see [6]). Following [7], we first review the definitions of classical indistinguishability obfuscation and then, following [1] and [8], we review the extension of these notions to the quantum setting. We recall that indistinguishability obfuscation comes in three variants: Computational, statistical, and perfect, in ascending order of strength. To define these, we first recall the following computational notions:

*Definition 5:*

1) For a classical Boolean circuit $F$, the *size* of the circuit is the total number of AND, OR, and NOT gates in $F$, which we denote by $|F|$.

2) A family $\mathcal{F} = \{C_n : n \in \mathbb{N}\}$ of classical Boolean circuits is called *polynomial-time uniform* if there is a classical deterministic Turing machine which, on input $1^n$, prints a description of $C_n$ in time poly$(n)$.
3) A *probabilistic polynomial-time algorithm* (PPT algorithm) is a polynomial-time uniform family $\mathcal{F} = \{C_n : n \in \mathbb{N}\}$ of classical Boolean circuits with the additional properties that: i) $C_n$ takes inputs from $\{0,1\}^n$ and ii) there exists a polynomial $p$ such that $C_n$ is augmented with a $p(n)$-bit coin register so that, for each $x \in \{0,1\}^n$, the output $C_n(x)$ is the probability distribution over the codomain of $C_n$ induced by uniformly sampling from $\{0,1\}^{p(n)}$.

Additionally, we recall the following statistical notions:

*Definition 6:* Given random variables $X, Y$ over a countable set $\Omega$, the *statistical distance* between $X$ and $Y$ is

$$\Delta(X,Y) := \frac{1}{2}\sum_{\omega\in\Omega} |\Pr[X=\omega] - \Pr[Y=\omega]|$$

Given a function $f : \mathbb{N} \to \mathbb{R}_{\geq 0}$, we say that $f$ is *negligible* if, for any integer $c > 0$, there exists an integer $N_c \geq 0$ such that, for all $n \geq N_c$, we have $f(n) \leq \frac{1}{x^c}$. That is, $f$ tends to zero faster than every inverse polynomial.

*Definition 7:* A *distribution ensemble* is a sequence $\mathcal{D} = \{D_n\}_{n\geq 1}$ of random variables, where, for each $n$, the domain of $D_n$ is countable. We usually take the domain of $D_n$ to be $\{0,1\}^{\ell(n)}$, where $\ell : \mathbb{N} \to \mathbb{N}$ is some function such that $\ell(n) = O(\text{poly}(n))$. Let $\mathcal{X} = \{X_n : n \in \mathbb{N}\}$ and $\mathcal{Y} = \{Y_n : n \in \mathbb{N}\}$ be distribution ensembles. Then:

1) We say that $\mathcal{X}, \mathcal{Y}$ are *perfectly* indistinguishable if, for all $n$, we have $\Delta(X_n, Y_n) = 0$.
2) We say that $\mathcal{X}, \mathcal{Y}$ are *statistically indistinguishable* if there exists a negligible function $f$ such that, for all $n$, $\Delta(X_n, Y_n) \leq f(n)$.
3) Lastly, we say that $\mathcal{X}, \mathcal{Y}$ are *computationally* indistinguishable if the following holds: For any PPT algorithm $\mathcal{A}$, where $\mathcal{A}$ takes as input $1^n$ and one sample $s$ from either $X_n$ or $Y_n$, and outputs either 0 or 1, there exists a negligible function $f$ such that, for all $n$,

$$|\Pr_{s\sim X_n}\left[\mathcal{A}(1^n, s) = 1\right] - \Pr_{s\sim Y_n}\left[\mathcal{A}(1^n, s) = 1\right]| \leq f(n)$$

Informally then, the last condition states that no PPT adversary, sampling from $X_n$ and $Y_n$, can succeed in distinguishing between $X_n$ and $Y_n$ with greater than negligible probability. With the above definitions in hand, we can define classical iO.

*Definition 8 (Classical Indistinguishability Obfuscation):* Let $\mathcal{F}$ be a family of classical Boolean circuits. For each $n \geq 1$, let $\mathcal{F}_n$ be the family of circuits of $\mathcal{F}$ with input length $n$. Let $\mathcal{O}$ be a algorithm that takes as input a circuit of $\mathcal{F}$ and outputs another classical Boolean circuit. We say that $\mathcal{O}$ is, respectively, a *efficient perfect/statistical/computational indistinguishability obfuscator (iO) for $\mathcal{F}$* if all of the following hold:

1) (Efficiency) $\mathcal{O}$ is a PPT algorithm.
2) (Preserving Functionality) For all $n \geq 1$, $F \in \mathcal{F}_n$, and $x \in \{0,1\}^n$, we have $(\mathcal{O}(F))(x) = F(x)$.
3) (Polynomial Slowdown) There exists a polynomial $p(n)$ such that, for all $n \geq 1$ and $F \in \mathcal{F}_n$, we have $|\mathcal{O}(F)| = O(p(|F|))$. That is, $\mathcal{O}$ only enlarges $F$ by at most a factor of $p(|F|)$.
4) (Indistinguishability) There exists an integer $N$ such that the following holds: Let $\{F_{1,n} : n \geq N\}$ and $\{F_{2,n} : n \geq N\}$ be two sequences of circuits in $\mathcal{F}$, where, for each $n \geq N$, $F_{1,n}, F_{2,n} \in \mathcal{F}_n$, and, in particular, $F_{1,n}$ and $F_{2,n}$ compute the same function and $|F_{1,n}| = |F_{2,n}|$. Then the distribution ensembles $\{\mathcal{O}(F_{1,n}) : n \geq N\}$ and $\{\mathcal{O}(F_{2,n}) : n \geq N\}$ are, respectively, perfectly/statistically/computationally indistinguishable.

Note that our functional equivalence requirement follows [1], rather than [7]. A different notion of functional equivalence is that $F$ and $\mathcal{O}(F)$ agree on all but a negligible number of inputs, that is, there exists a negligible function $g$ such that, for any $n$ and any $F \in \mathcal{F}_n$,

$$\Pr\left[x \in \{0,1\}^n : F(x) \neq (\mathcal{O}(F))(x)\right] \leq g(n) \tag{1}$$

where the probability is taken over $\mathcal{O}$'s coins. In this paper, we will always work with the *exact* functional equivalence definition. To pass to the quantum setting, we first recall the quantum analogues to Definition 5:

*Definition 9:*

1) A *generalized quantum circuit* $C$ on $n$ input qubits consists of a quantum circuit on $n$ input qubits which is augmented with trace-out and measurement operations. The *size* of the circuit is its gate count, denoted by $|C|$.
2) Let $\mathcal{F} = \{C_n : n \in \mathbb{N}\}$ be a family of generalized quantum circuits. We say that $\mathcal{F}$ is *polynomial-time uniform* if there is a classical deterministic Turing machine which, on input $1^n$, prints a classical description of $C_n$ in time poly$(n)$.
3) A *quantum polynomial-time algorithm* is a polynomial-time uniform family $\{C_n : n \in \mathbb{N}\}$ of generalized quantum circuits with the additional property that, for each $n \in \mathbb{N}$, $C_n$ takes $n$ qubits as input.

Again following [1], we also define what it means for an obfuscator to be quantum-secure:

*Definition 10:* Let $\mathcal{F}$ and $\mathcal{O}$ be as in Definition 8 and let $\mathcal{O}$ be a computational indistinguishability obfuscator for $\mathcal{F}$. We say that $\mathcal{O}$ is *quantum-secure* if, in Definition 7, the PPT algorithm $\mathcal{A}$ is replaced by a quantum polynomial-time algorithm as adversary.

Note that the modifier "quantum-secure" only affects $\mathcal{O}$ in the case where $\mathcal{O}$ is a computational indistinguishability obfuscator. In the statistical/perfect cases, additional computational power conferred to an adversary does not make a difference. We now review some known results about classical indistinguishability obfuscation. We have the following, from [7], that there is a polynomial-sized family of of Boolean circuits for which perfect/statistical indistinguishability obfuscation is not possible even if the efficiency requirement is dropped.

*Theorem 1 (Goldwasser, Rothblum [7]):* If the family of 3-CNF formulas admits a (not necessarily efficient) statistical indisitnguishability obfuscator, then the polynomial hierarchy collapses to the second level.

Of course, this also implies that perfect or statistical iO is (probably) not possible when using the modified functionality-preserving requirement in Definition 8, as this requirement is stronger than condition (1). Thus, Theorem 1 leaves computational iO as the best we can hope for without restricting the circuit domain. In 2021, Jain, Lin, and Sahai showed that computational iO for families of polynomial-sized circuits is possible under certain (well-founded) assumptions. However, the procedure in [9] is not quantum-secure and, to our knowledge, the challenge of constructing a *quantum-secure* efficient classical computational iO remains open. Now, following [1] and [8], we extend the notions in Definition 8 to the quantum setting. The review of the gate-teleportation protocol of [1] in Section III makes explicit why the algorithm in Definition 11 outputs pairs $(|\psi\rangle, F)$. Informally, $|\psi\rangle$ can be thought of as a "resource state" which the end user, holding onto $\mathcal{O}(F)$, can use to appy $F$ to an arbitrary state without learning about the circuit description of $F$.

*Definition 11 (Quantum Indistinguishability Obfuscation, [1]):* Let $\mathcal{F}$ be a polynomial-time uniform family of reversible quantum circuits. For each $n \geq 1$, let $\mathcal{F}_n$ be the family of $n$-qubit circuits of $\mathcal{F}$. Let $\mathcal{O}$ be an algorithm that takes as input a circuit $F$ of $\mathcal{F}$ and outputs a *pair* $(|\psi\rangle, F')$, where $F'(|\psi\rangle, \cdot)$ is a quantum circuit. We say that $\mathcal{O}$ is, respectively, a *efficient computational/statistical/perfect quantum indistinguishability obfuscator (qiO) for* $\mathcal{F}$ if there exists a function $\ell : \mathcal{F} \to \mathbb{N}$ such that all of the following hold:

1) (Efficiency) $\mathcal{O}$ is a quantum polynomial-time algorithm.
2) (Preserving Functionality) For all $n \geq 1$ and $F \in \mathcal{F}_n$, letting $(|\psi\rangle, F') = \mathcal{O}(F)$, the circuits $F'(\psi, \cdot)$ and $F(\cdot)$ are perfectively equivalent, where $|\psi\rangle$ is an $\ell(F)$-qubit state and $F'$ is a circuit of type $(\ell(F) + n, n)$.
3) (Polynomial Slowdown) There exists a polynomial $p(n)$ such that, for any $F \in \mathcal{F}$, letting $\mathcal{O}(F) = (|\phi\rangle, F')$. we have both $\ell(F) \leq p(|F|)$ and $|F'| \leq p(|F|)$.
4) (Indistinguishability) There exists an integer $N$ such that the following holds: Let $\{F_{1,n} : n \geq N\}$ and $\{F_{2,n} : n \geq N\}$ be two sequences of circuits, where, for each $n \geq N$, $F_{1,n}$ and $F_{2,n}$ are perfectly equivalent circuits of $\mathcal{F}_n$ with $|F_{1,n}| = |F_{2,n}|$. Then the distribution ensembles $\{\mathcal{O}(F_{1,n}) : n \geq N\}$ and $\{\mathcal{O}(F_{2,n}) : n \geq N\}$ are, respectively, perfectly/statistically/computationally indistinguishable.

As above, for the case of a computational qiO, we say that $\mathcal{O}$ is *quantum-secure* if computational indistinguishability holds against a quantum polynomial-time adversary. Also, for the functional equivalence requirement, we are again following [1], rather than [8]. In the latter paper, the equivalence condition corresponding to 2) only requires that $\|F(\cdot) - F'(|\psi\rangle, \cdot)\|_\diamond \leq f(n)$ for some negligible function $f$. As noted in [1], there is a very limited class of quantum circuits for which efficient perfect indistinguishability obfuscation is possible (in the sense of Definition 11), namely, the Clifford circuits.

*Theorem 2:* The class of Clifford circuits admits an efficient perfect qiO.

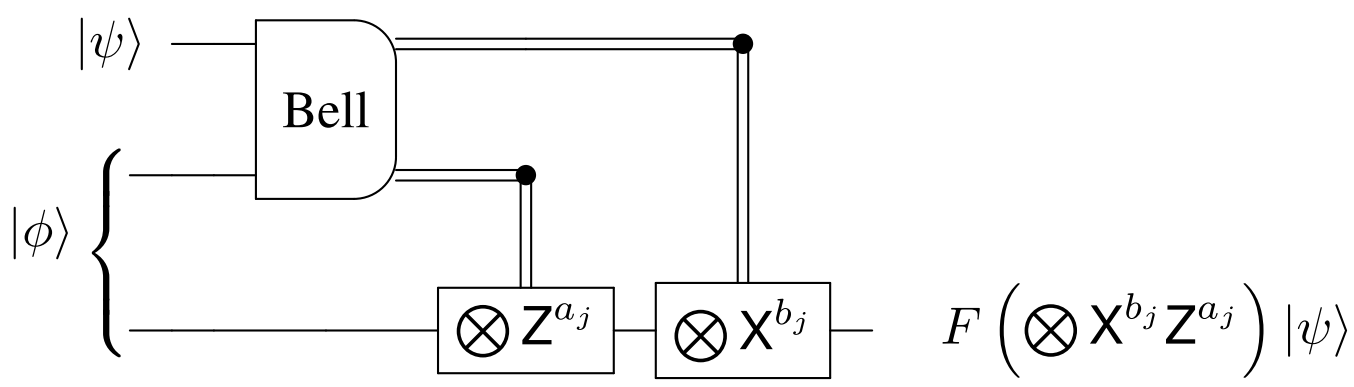

Fig. 1. The teleportation gadget acting on the $3n$-qubit system $|\psi\rangle|\phi\rangle$.

This holds due to the well-known fact (see [10] and [11]), that every Clifford circuit admits a *canonical form*. More precisely:

*Theorem 3:* Given as input an $n$-qubit Clifford circuit $C$, we can, in time poly$(n, |C|)$, output a canonical Clifford circuit $C'$, with $O(n^2)$ gates, such that $C$ and $C'$ instantiate the (exact) same unitary.

## III. Gate Teleportation and Conjugate Encoding

We begin by recalling the usual procedure for teleporting an $n$-qubit state by making use of the Bell basis:

*Definition 12:* The four 2-qubit *Bell states* are given by

$$|\beta_{00}\rangle = \frac{|00\rangle+|11\rangle}{\sqrt{2}} \quad |\beta_{01}\rangle = \frac{|01\rangle+|10\rangle}{\sqrt{2}}$$

$$|\beta_{10}\rangle = \frac{|00\rangle-|11\rangle}{\sqrt{2}} \quad |\beta_{11}\rangle = \frac{|01\rangle-|10\rangle}{\sqrt{2}}$$

For each binary string $s = (a_1, \ldots, a_n, b_1, \ldots, b_n) \in \mathbb{F}_2^{2n}$, we define an associated $n$-qubit (generalized) *Bell state*:

$$|\beta_s\rangle = \bigotimes_{j=1}^{n} |\beta_{a_j b_j}\rangle$$

The set $\{|\beta_s\rangle : s \in \mathbb{F}_2^{2n}\}$ forms an orthonormal basis for the Hilbert space $(\mathbb{C}^2)^{\otimes 2n}$ so, for any $2n$-qubit state $|\sigma\rangle$, there is a unique expression

$$|\sigma\rangle = \sum_{s \in \mathbb{F}_2^{2n}} \alpha_s |\beta_s\rangle \text{ where } \sum_{s \in \mathbb{F}_2^{2n}} |\alpha_s|^2 = 1$$

A *Bell measurement* on $|\sigma\rangle$ outputs a string of $\mathbb{F}_2^{2n}$ according to the probability distribution where $s$ is obtained with probability $|\alpha_s|^2$, and obtaining this outcome leaves the quantum system in state $|\beta_s\rangle$.

With this definition in hand, we can define our teleportation procedure. Suppose we have an $n$-qubit circuit $F$ that we want to apply to an arbitrary $n$-qubit state $|\psi\rangle$. Instead of applying $F$ to $|\psi\rangle$ directly, we can use the $3n$-qubit teleportation protocol to obtain the state $F|\psi\rangle$ only by applying $F$ to one half of a $2n$-qubit Bell state, by performing the following:

1) Let $|\phi\rangle := (F \otimes \mathsf{I}^{\otimes n})|\beta_{\mathbf{0}}\rangle$, where $\mathbf{0}$ denotes the all-zero vector of $\mathbb{F}_2^{2n}$.
2) Perform a Bell measurement on the first $2n$ qubits of $|\psi\rangle|\phi\rangle$ to obtain outcome $x = (a_1, \ldots, a_n, b_1, \ldots, b_n) \in \mathbb{F}_2^{2n}$.
3) Obtain a state on the last $n$ registers which is $F\mathsf{P}^x|\psi\rangle$ up to global phase, as illustrated in Figure 1.

After performing the teleportation, we recover $F|\psi\rangle$ from $F\mathsf{P}^x|\psi\rangle$ by applying the correction unitary $F_{\text{corr}} := F\mathsf{P}^x F^\dagger$. Then, up to global phase, we have

$$F_{\text{corr}}F\left(\bigotimes_{j=1}^{n}\mathsf{X}^{b_j}\mathsf{Z}^{a_j}\right)|\psi\rangle = F|\psi\rangle$$

Informally, the gate-teleportation protocol of [1] offloads the task of obfuscating the quantum circuit $F$ onto the classical task of obfuscating a classical circuit $f$ which, on input $x \in \mathbb{F}_2^{2n}$, outputs a description of the unitary $F\mathsf{P}^xF^\dagger$, where this classical description $f(x)$ reveals nothing about $F\mathsf{P}^xF^\dagger$ except for its functionality. We make this precise below. We first recall some information about certain gate sequences expressed as linear combinations of Paulis.

$$\mathsf{S} = \left(\frac{1+i}{2}\right)\mathsf{I} + \left(\frac{1-i}{2}\right)\mathsf{Z} = \frac{1}{\sqrt{2}}\left(e^{i\pi/4}\mathsf{I} + e^{-i\pi/4}\mathsf{Z}\right)$$

We then have $e^{-i\pi/4}\mathsf{S} = \frac{\mathsf{I}+i\mathsf{Z}}{\sqrt{2}}$. The following identity is then straightforward to check by following the action of each side on a computational basis state, and using the identity $\mathsf{SX} = i\mathsf{XZS}$.

*Observation 1:* Let $u = (a, b) \in \mathbb{F}_2^2$ and let $u' = (a \oplus b, b)$. Then

$$\mathsf{T}\mathsf{P}^u\mathsf{T}^\dagger = e^{-bi\pi/4}\mathsf{S}^b\mathsf{P}^u = \begin{cases}\mathsf{P}^u \text{ if } b = 0\\ \left(\frac{\mathsf{I}+i\mathsf{Z}}{\sqrt{2}}\right)\mathsf{P}^u \text{ if } b = 1\end{cases}$$
$$= \frac{\mathsf{P}^u + (-1)^{b(a\oplus 1)}\mathsf{P}^{u'}}{2^{1-\frac{b}{2}}}$$

Additionally, we recall the following usual notation.

*Definition 13:* For $n$-qubit quantum circuits $C_1, C_2$, we write $C_1 \equiv C_2$ to mean that there is a $\theta \in [0, 2\pi)$ such that $C_1C_2^\dagger = e^{i\theta}\mathsf{I}^{\otimes n}$. Otherwise, we write $C_1 \not\equiv C_2$.

We first note that the set of $n$-qubit Paulis of order two is precisely $\{-\mathsf{I}^{\otimes n}\} \cup \{\pm\mathsf{P}^x : x \in \mathbb{F}_2^{2n} \setminus \{\mathbf{0}\}\}$. Since conjugation preserves order, for any $n$-qubit Clifford $C$ and any $y \in \mathbb{F}_2^{2n}\setminus\{\mathbf{0}\}$, there is a $\tau \in \mathbb{F}_2$ and a $z \in \mathbb{F}_2^{2n}\setminus\{\mathbf{0}\}$ such that $C\mathsf{P}^yC^\dagger = (-1)^\tau\mathsf{P}^z$. More precisely, we have the following.

*Proposition 1* (Clifford Update Rules): Let $((a, b), \tau) \in \mathbb{F}_2^2 \times \mathbb{F}_2$, corresponding to the single-qubit Pauli $(-1)^\tau i^{ab}\mathsf{X}^b\mathsf{Z}^a$. Then the action by conjugation of each $\mathsf{H}, \mathsf{S}, \mathsf{Z}, \mathsf{X}$ on this Pauli corresponds to the following respective maps $\mathbb{F}_2^2 \times \mathbb{F}_2 \to \mathbb{F}_2^2 \times \mathbb{F}_2$.

$$\mathsf{H} : ((a, b), \tau) \to ((b, a), \tau \oplus a \cdot b)$$

$$\mathsf{S} : ((a, b), \tau) \to ((a \oplus b, b), \tau \oplus a \cdot b)$$

$$\mathsf{Z} : ((a, b), \tau) \to ((a, b), \tau \oplus b)$$

$$\mathsf{X} : ((a, b), \tau) \to ((a, b), \tau \oplus a)$$

For $((a, b), (c, d), \tau) \in \mathbb{F}_2^4 \times \mathbb{F}_2$, the action by conjugation of $\mathsf{CZ}$ on the corresponding two-qubit Pauli $(-1)^\tau i^{ab+cd}\mathsf{X}^b\mathsf{Z}^a \otimes \mathsf{X}^d\mathsf{Z}^c$ is given by the following map $\mathbb{F}_2^4 \times \mathbb{F}_2 \to \mathbb{F}_2^4 \times \mathbb{F}_2$.

$$\mathsf{CZ} : ((a, b), (c, d), \tau)$$
$$\to ((a \oplus d, b), (c \oplus b, d)), \tau \oplus bd(a \oplus c))$$

These update rules immediately give us the following.

*Proposition 2:* Let $C$ be an $n$-qubit Clifford circuit with gate description $C = g_m \cdots g_1$. Then there is a classical Boolean circuit $f : \{0, 1\}^{2n} \to \{0, 1\}^{2n} \times \{0, 1\}$ which, on input $a \in \{0, 1\}^{2n}$, outputs $(b, k)$ where $C\mathsf{P}^aC^\dagger = (-1)^k\mathsf{P}^b$. Furthermore, $f$ can be compiled in time $O(nm)$, and runs in time $O(nm)$.

More generally, the Clifford update rules, together with Observation 1, yield the following:

*Proposition 3:* There is a classical deterministic circuit $f$, which, given as input a pair $(C, x)$, where $C \in \mathcal{T}_n$ and $x \in \mathbb{F}_2^{2n}$, outputs the sequence $\left(x_1, \langle C\mathsf{P}^xC^\dagger, \mathsf{P}^{x_1}\rangle\right), \cdots, \left(x_k, \langle C\mathsf{P}^xC^\dagger, \mathsf{P}^{x_k}\rangle\right)$, where

i) $\mathsf{P}^{x_1}, \cdots, \mathsf{P}^{x_k}$ are precisely the Paulis with nonzero coefficient in the Pauli expansion of $C\mathsf{P}^xC^\dagger$ (under some suitable encoding), where $x_1, \cdots, x_k$ are in lexicographical order; AND
ii) If $C$ has $t$ $\mathsf{T}$-gates, and $m$ Clifford gates then $f(C, x)$ is outputted in time $\text{poly}(m, n, t) \cdot 2^{\min\{t,2n\}}$.

We generalize this notion in Proposition 3 below.

*Definition 14:* Let $\mathcal{F}$ be a family of reversible quantum circuits with gates drawn from some fixed, finite gate set. For each $n \geq 1$, let $\mathcal{F}_n$ be the set of $n$-qubit circuits in $\mathcal{F}$.

1) We define a *conjugate-encoding* for $\mathcal{F}$ to be a classical, deterministic Turing machine $\mathcal{A} : \bigcup_{n\geq 1}\mathcal{F}_n \times \mathbb{F}_2^{2n} \to \{0, 1\}^*$ such that, for any $n \geq 1$ and any $(C, x), (C', x') \in \mathcal{F}_n \times \mathbb{F}_2^{2n}$, if $C\mathsf{P}^xC^\dagger$ and $C'\mathsf{P}^{x'}(C')^\dagger$ are distinct unitaries, then $\mathcal{A}(C, x) \neq \mathcal{A}(C', x')$. We think of the output of $\mathcal{A}$ on input $(C, x)$ as some suitable binary encoding of the unitary $C\mathsf{P}^xC^\dagger$.
2) Given such an $\mathcal{A}$, we define an *$\mathcal{A}$-recovery algorithm* $\mathcal{R}$ to be a classical deterministic Turing machine accepting inputs of the form $\mathcal{A}(C, x)$, where, on this input, $\mathcal{R}$ outputs a gate description of the unitary $C\mathsf{P}^xC^\dagger$. For $\mathcal{F}' \subseteq \mathcal{F}$, the pair $(\mathcal{A}, \mathcal{R})$ is said to be *efficient* over $\mathcal{F}'$ if:
   a) On input $(C, x)$, for an $n$-qubit $C \in \mathcal{F}'$, the output of $\mathcal{A}$ is produced in time polynomial in both $n, |C|$; AND
   b) On input $\mathcal{A}(C, x)$, the output $\mathcal{R}(\mathcal{A}(C, x))$ is produced in time polynomial in $n$ and $|\mathcal{A}(C, x)|$, where $|\mathcal{A}(C, x)|$ is the length of the output $\mathcal{A}$ on $(C, x)$.

We call $\mathcal{F}$ the *circuit-domain* of $\mathcal{A}$. We say that $\mathcal{A}$ is *function-determined* over $\mathcal{F}$ if it satisfies the following stronger converse to 1) of Definition 14: For any $n \geq 1$ and $C, C' \in \mathcal{F}_n$, if $C \equiv C'$, then $\mathcal{A}(C, x) = \mathcal{A}(C', x)$ for each $x \in \mathbb{F}_2^{2n}$.

For example, given gate description $C = g_m \cdots g_1$, a trivial $\mathcal{A}$ simply consists of outputting the gate sequence $g_m \cdots g_1\mathsf{P}^xg_1^\dagger \cdots g_m^\dagger$, but this reveals the complete gate description of $C$ and is not function-determined. Conversely, a trivial (but inefficient) function-determined conjugate encoding would be to perform matrix multiplication to output the $2^n \times 2^n$-unitary $C\mathsf{P}^xC^\dagger$. The algorithm $f$ in Proposition 3 is also a function-determined conjugate-encoding, since the Paulis form a basis for $M_{2^n \times 2^n}(\mathbb{C})$. Of course, $f$ is not efficient in general, since, for $n$ input registers, $f(C, x)$ is potentially a sequence of $4^n$ terms. The protocol of [1] yields the more general protocol below that relates conjugate-encodings to qiOs over the specified circuit domain.

*Theorem 4:* Let $\mathcal{F}$ be a family of Clifford+T circuits and let $\mathcal{A}$ be a function-determined conjugate-encoding with circuit domain $\mathcal{F}$. Suppose there is an $\mathcal{A}$-recovery algorithm $\mathcal{R}$ such that $(\mathcal{A}, \mathcal{R})$ is efficient over $\mathcal{F}$. Then, if there is a quantum-secure classical computational iO, there is also a quantum-secure computational qiO for $\mathcal{F}$. An appropriate modification of the gate-teleportation protocol from [1] provides the qiO for $\mathcal{F}$.

*Proof:* Let $F \in \mathcal{F}$ be a fixed input to the qiO, where $F$ has $n$ qubits. Given this $F$, we can regard $\mathcal{A}(F, \cdot)$ as a classical circuit $\mathbb{F}_2^{2n} \to \{0, 1\}^*$. Let $\mathcal{A}'_F$ be the classical circuit obtained by applying the classical computational iO to $\mathcal{A}(F, \cdot)$. The output of the qiO is a pair $(F', |\phi\rangle)$, where $|\phi\rangle$ is prepared in the following way:

1) Prepare the $n$-qubit Bell state $|\beta_{\mathbf{0}}\rangle$.
2) Apply $F$ to the rightmost $n$-qubits of $|\beta_{\mathbf{0}}\rangle$ to obtain the $2n$-qubit system $|\phi\rangle := (\mathsf{I}^{\otimes n} \otimes F)|\beta_{\mathbf{0}}\rangle$.

and $F'$ is the circuit which, on arbitrary input an $n$-qubit $|\psi\rangle$, consists of doing the following:

1) Perform a Bell measurement on the first $2n$ qubits of the $3n$-qubit system $|\phi\rangle|\psi\rangle$ to obtain measurement outcome $s \in \mathbb{F}_2^{2n}$. This leaves the last $n$ registers in the state $F\mathsf{P}^s|\psi\rangle$ up to global phase.
2) Apply $\mathcal{A}'_F$ to the outcome $s$, and then apply $\mathcal{R}$ to $\mathcal{A}'_F(s)$.
3) Apply the resulting gate sequence $\mathcal{R}(\mathcal{A}'_F(s))$ to $F\mathsf{P}^s|\psi\rangle$ to obtain $F|\psi\rangle$ up to global phase. $\square$

Due to the run-time of Proposition 3, the protocol of Theorem 4 is efficient for circuits with logarithmically many T-gates.

## IV. The Main Results and an Overview of the Proofs

As a first task, we prove that function-determined conjugate-encoding is possible for circuits of T-depth $\leq 1$. This is, in a certain sense, the simplest natural class of Clifford+T circuits which contains some circuits with a superlogarithmic T-count. In fact, our observations above in Sections III let us do this immediately:

*Theorem 5:* There is a conjugate encoding $\mathcal{A}$ and an $\mathcal{A}$-recovery $\mathcal{R}$ such that $\mathcal{A}$ has circuit-domain $\mathcal{T}^1$, and $(\mathcal{A}, \mathcal{R})$ is efficient over $\mathcal{T}^1$. Thus: Efficient quantum-secure computational indistinguishability obfuscation is possible for this circuit class if there is an efficient quantum-secure classical computational iO for the domain of all classical Boolean circuits.

*Proof:* Given $(C, x) \in \mathcal{T}_n^1 \times \mathbb{F}_2^{2n}$, we note that, by Observation 1, $C\mathsf{P}^x C^\dagger$ is a Clifford unitary. That is, $C$ lies in the third level of the the Clifford hierarchy, it maps Paulis to Cliffords under conjugation (see [12] for more details on the Clifford hierarchy). We can easily convert the gate description $C\mathsf{P}^x C^\dagger$ into a sequence of $O(|C|)$ Clifford gates, again using Observation 1. Our $(\mathcal{A}, \mathcal{R})$ then consists of taking $(C, x)$ and outputting the canonical Clifford gate description of $C\mathsf{P}^x C^\dagger$ as in Theorem 3. $\square$

For this very restricted class of circuits, we thus obtain an immediate improvement over Proposition 3, because the Pauli expansion of $C\mathsf{P}^x C^\dagger$ for $(C, x) \in \mathcal{T}_n^1 \times \mathbb{F}_2^{2n}$ can have up to $2^n$ nonzero terms. The ability to produce a function-determined conjugate encoding for a given class of circuits is closely related to the problem of determining whether a given circuit instantiates the identity exactly up to global phase (we make this connection explicit in Section VI). We recall first the usual non-identity check decision problem for quantum circuits.

*Definition 15 (Non-Identity Check, [13]):* We are given as input a classical description of a reversible quantum circuit $U$ on $n$ qubits and two real numbers $a, b$ with $b - a \geq 1/\text{poly}(n)$, where it is promised that $\min_{\psi \in [0, 2\pi)} \|U - e^{i\psi}\mathsf{I}^{\otimes n}\|$ is either $> b$ or $< a$. The problem Non-Identity Check (NIC) is to decide which of these holds.

Analogous to this, we define the following.

*Definition 16:* The decision problem ENIC (*Exact Non-Identity Check*) is defined as follows: Given as input a classical description of an $n$-qubit quantum circuit $C$, decide if there exists a $\theta \in [0, 2\pi)$ such that $C = e^{i\theta}\mathsf{I}^{\otimes n}$. In particular, $C$ is a yes-instance of ENIC if $C \not\cong \mathsf{I}^{\otimes n}$.

We recall that the complexity classes NQP and QMA are, informally speaking, different quantum analogues of NP. In particular, NP $\subseteq$ NQP $\cap$ QMA (see [14] and [15] respectively for more details on the classes NQP and QMA). The problem of ENIC was introduced and studied in [16], where Tanaka established that ENIC is NQP-hard, analogous to fact that NIC is QMA-hard as shown in [17]. Indeed, ENIC was introduced precisely in order to provide an example of a complete problem for the class NQP. The non-exact version, NIC, is QMA-complete even for circuits of constant depth, as shown in [13]. A more detailed discussion of QMA and NQP is not essential for our purposes. The remainder of this paper focuses solely on classical hardness, and the only complexity classes we are concerned with are P, NP, and #P. We now relate ENIC to some problems that are more suitable to our setting.

*Definition 17:*

1) We define CONJUGATE to be the following functional problem: On input a classical description of an $n$-qubit quantum circuit $C$ and an $x \in \mathbb{F}_2^{2n}$, output the exact value of $\langle C\mathsf{P}^x C^\dagger, \mathsf{P}^x\rangle$.
2) We define SUPPORT to be the following decision problem. On input a classical description of an $n$-qubit quantum circuit $C$ and an $x \in \mathbb{F}_2^{2n}$, decide whether or not $\mathsf{P}^x$ appears with nonzero coefficient in the Pauli-expansion of $C\mathsf{P}^x C^\dagger$.
3) Likewise, we define COMMUTE to be the following decision problem. On input a classical description of an $n$-qubit quantum circuit $C$ and an $x \in \mathbb{F}_2^{2n}$, decide whether or not $C\mathsf{P}^x C^\dagger = \mathsf{P}^x$.

Our main result for this paper is the following:

*Theorem 6:* All of the following hold.

M1) CONJUGATE is #P-hard over domain $\mathcal{T}^3$.

M2) ENIC over domain $\mathcal{T}^2$ lies in P, but, conversely, there exists an absolute constant $K > 0$ such that all three of the decision problems ENIC and SUPPORT and COMMUTE are NP-hard over the domain $\mathcal{T}^{K\log(n)}$.

M3) If P ≠ NP, then, with $K$ as above, the class $\mathcal{T}^{K\log(n)}$ does not admit efficient conjugate encoding, i.e. the protocol of [1] cannot be extended to an efficient computational qiO for this circuit class.

The definition of ENIC used in Definition 16 is chosen to be analogous to Definition 15, but it is slightly different to that of [16] (in [16], we are trying to determine whether $C = \mathsf{I}^{\otimes n}$ exactly, not just up to global phase), but Theorem 6 implies that both versions of the problem are NP-hard for logarithmic-T-depth quantum circuits in any case. In particular, any instance $(C, x)$ of COMMUTE is equivalent to deciding whether the circuit $C\mathsf{P}^xC^\dagger\mathsf{P}^x$ exactly instantiates the unitary $\mathsf{I}^{\otimes n}$. Definition 16 is more natural for our purposes because conjugation destroys global phases.

The limits of the protocol of [1] are of interest since, as noted in [18], the protocol of [1] is, at the time of writing, the only known way to obtain indistinguishability obfuscation for expressive classes of quantum circuits from quantum-secure indistinguishability obfuscation of *classical* circuits. Theorem 6 (probably) rules out ENIC for circuits of low T-depth, as well as ruling out extending the protocol of [1] to this class of circuits. However, it is not true in general that the existence of an arbitrary efficient computational qiO $\mathcal{O}$ (in the sense of Definition 11) for a given circuit class $\mathcal{F}$ necessarily implies that it is possible to efficiently decide whether two elements of $\mathcal{F}$ instantiate the same unitary (unless $\mathcal{O}$ is deterministic).

We prove the main result by a sequence of reductions where we relate the stated problems in Theorem 6 to some hardness results for codeword weights in binary linear codes. Below, we briefly describe the intuition behind this link. Recall that the *Hamming weight* of a $v \in \mathbb{F}_2^n$ is the number of nonzero entries of $v$. Suppose we begin with a single $n$-qubit Pauli $\mathsf{P}^x$, where $x = (a, b)$ for $a, b \in \mathbb{F}_2^n$. Suppose now that we conjugate $\mathsf{P}^x$ by $\mathsf{T}^{\otimes n}$. In that case, we obtain a Hermitian Clifford $C_1$ whose Pauli expansion is a uniform-magnitude linear combination of Paulis, where the support (the set of Paulis with nonzero coefficient) corresponds to an affine subspace $A \subseteq \mathbb{F}_2^{2n}$ of dimension $|b|$, and the common magnitude is $\left(1/\sqrt{2}\right)^{|b|}$. More precisely, using Observation 1, we get

$$C_1 = \mathsf{T}^{\otimes n}\mathsf{P}^x(\mathsf{T}^{\otimes n})^\dagger = \left(\frac{1}{\sqrt{2}}\right)^{|b|}\left(\prod_{\substack{1\le j\le n\\ b_j=1}}(\mathsf{I} + i\mathsf{Z}_j)\right)\mathsf{P}^x \tag{2}$$

In particular, the underlying vector space $A \oplus x$ is generated by an ordered basis $\mathbf{u}$ consisting of $|b|$ standard Z-basis vectors, each of which corresponds to a Pauli that anticommutes with $\mathsf{P}^x$. The order of the factors in the product term in the RHS of (2) does not matter, as the factors are commutative. If we conjugate the expression in (2) by a Clifford, then we map $\mathsf{P}^x$ to another Pauli $\mathsf{P}^{x'}$ and map $\mathbf{u}$ to an ordered basis $\mathbf{u}'$ of a different subspace of $\mathbb{F}_2^{2n}$, where, again, this subspace consists of pairwise-commuting Paulis, and each Pauli corresponding to a basis vector in $\mathbf{u}'$ anticommutes with $\mathsf{P}^{x'}$. These observations form the starting point for how we later describe the Pauli coefficients of arbitrary PC-circuits.

If, instead of just conjugating by a Clifford, we conjugate $C_1$ by a T-depth-one circuit to produce a new PC-circuit $C_2$, each individual term in the initial Pauli expansion now itself expands into an affine space of Pauli terms. We can regard $C_2$ as a uniform-magnitude linear combination of conjugates of the $A$-Paulis by the same T-depth-one circuit. Under some conditions on the phases which we make precise later, we get that, for each $v \in \mathbb{F}_2^{2n}$,

$$\langle C_2, \mathsf{P}^v\rangle = \left(\frac{1}{\sqrt{2}}\right)^{|b|+\dim(X_v)} |A \cap X_v|$$

where $X_v \subseteq \mathbb{F}_2^{2n}$ is an affine space depending on $v$. That is, we can construct $C_2$ so that each coefficient of $C_2$ is a uniform average of the $C_1$-coefficients as we run over $X_v$. In particular, since $A \cap X_v$ is itself affine, each nonzero coefficient in $C_2$ is again an integer power of $1/\sqrt{2}$, although now they are no longer all the same power, nor is the support necessarily an affine space anymore. If we conjugate one more time by a T-depth-one circuit to obtain a new PC-circuit $C_3$, then we can regard $C_3$ as a uniform-magnitude linear combination of conjugates of the$A$-Paulis by the same T-depth-two circuit. With some careful construction of $C_3$, we can find a $v \in \mathbb{F}_2^{2n}$ satisfying

$$\langle C_3, \mathsf{P}^v\rangle = \left(\frac{1}{\sqrt{2}}\right)^{\dim(Y_v)} \sum_{u\in Y_v}\left(\frac{1}{\sqrt{2}}\right)^{|u|}$$

where $Y_v \subseteq \mathbb{F}_2^{2n}$ is an affine space depending on $v$. This illustrates the intuitive link between these coefficients and the weight enumerator polynomials of binary linear codes evaluated at the point $1/\sqrt{2}$, and why each round of conjugation by a tensor product of simultaneous T-gates greatly increases the difficulty of computing the Pauli coefficients of PC-circuits.

The remainder of this paper is structured as follows:

1) In Section V, we review some additional notions and facts about the action of Cliffords on subspaces of $\mathbb{F}_2^{2n}$. In particular, as indicated above, we will need some basic notions and facts about subspaces of $\mathbb{F}_2^{2n}$ whose corresponding Paulis pairwise commute.
2) In Section VI, we we prove that both ENIC and COMMUTE for circuits of T-depth $2d + O(\log(n))$ are at least as hard as SUPPORT for circuits for T-depth $d$.
3) In Section VII, we show that we can use the Pauli-coefficients of low-T-depth PC-circuits to encode arithmetic operations, i.e. certain linear combinations Pauli coefficients of PC-circuits can themselves be encoded as the Pauli coefficients of larger PC-circuits without adding too much T-depth-complexity (and only polynomial blow-up in the number of qubits).
4) In Sections VIII-IX, we show that, given a pair $(C, x) \in \mathcal{T}_n^d \times \mathbb{F}_2^{2n}$, the Pauli expansion of $C\mathsf{P}^xC^\dagger$ can be written as a signed sum of powers of $1/\sqrt{2}$, indexed by a subset of $\mathbb{F}_2^{2nd}$. This will allow us to relate these linear combinations to weight enumerator polynomials of binary linear codes.

5) In Section X, we review some coding theory preliminaries and prove some additional coding theory facts that we need later. In particular, we review the known hardness results which we reduce to the decision problems in Theorem 6.
6) In Section XI, we complete the proof of Theorem 6. We show that, under certain conditions, for a binary linear code $V$ with weight enumerator polynomial $\mathrm{wt}_V(x)$, the following hold:
   i) We can construct a conjugate of a single Pauli by a $\mathsf{T}$-depth three circuit, where one of the Pauli coefficients of the resulting PC-circuit encodes a term of the form $\mathrm{wt}_V(1/\sqrt{2})$, which is #P-hard to compute in the worst case.
   ii) Using the work of Section VII, we can construct a poly($n$)-qubit PC-circuits of $\mathsf{T}$-depth $O(\log(n))$ for which one of the Pauli coefficients encodes a certain linear combinations of terms of the form $\mathrm{wt}_V(2^{-k/2})$, where it is NP-hard to decide in the worst case whether this linear combination evaluates to zero.

## V. Cliffords as Symplectomorphisms

We first recall some notions from linear algebra. Throughout this paper, we write $\mathbf{0}_m$ and $\mathbf{1}_m$ to denote the respective all-zero and all-one vector of $\mathbb{F}_2^m$. When the underlying dimension is clear, we just write $\mathbf{0}$ and $\mathbf{1}$ respectively. The labelling in Definition 4 gives an isomorphism between $\mathbb{F}_2^{2n}$ and the phaseless Pauli group $\mathcal{P}_n/\langle \pm 1, \pm i\rangle$. The vector space $\mathbb{F}_2^{2n}$ is equipped with a symplectic bilinear form $B : \mathbb{F}_2^{2n} \times \mathbb{F}_2^{2n} \to \mathbb{F}_2$, where

$$B((a_1,\ldots,a_n,b_1,\cdots,b_n),(a_1',\ldots,a_n',b_1',\cdots,`b_n')) \equiv \sum_{j=1}^{n}(a_jb_j' + a_j'b_j) \bmod 2$$

For $v, v' \in \mathbb{F}_2^{2n}$, the Paulis $\mathsf{P}^v$ and $\mathsf{P}^{v'}$ commute if and only if $B(v, v') = 0$. Otherwise they anticommute. A subspace of $\mathbb{F}_2^{2n}$ is called *isotropic* if the bilinear form $B$ vanishes over $V$. Thus, the isotropic subspaces of $\mathbb{F}_2^{2n}$ are precisely those subspaces $V$ such that the elements of $\{\mathsf{P}^v : v \in V\}$ mutually commute. We also define the following:

*Definition 18:* For each $1 \le j \le n$, we let $e_{\mathsf{Z}}^j \in \mathbb{F}_2^{2n}$ denote the $j$th standard basis vector in $\mathbb{F}_2^{2n}$ and, likewise, $e_{\mathsf{X}}^j$ denote the $(j+n)$th standard basis vector. For any $1 \le r \le n$, we let $\mathbf{E}_{\mathsf{Z}}^r$ and $\mathbf{E}_{\mathsf{X}}^r$ denote the respective ordered bases $(e_{\mathsf{Z}}^1,\ldots,e_{\mathsf{Z}}^r)$ and $(e_{\mathsf{X}}^1,\ldots,e_{\mathsf{X}}^r)$. We call a subspace $V$ of $\mathbb{F}_2^{2n}$ a $\mathsf{Z}$-*subspace* (resp. $\mathsf{X}$-*subspace*) if $V \subseteq \mathrm{span}(\mathbf{E}_{\mathsf{Z}}^n)$ (resp. $V \subseteq \mathrm{Span}(\mathbf{E}_{\mathsf{X}}^n)$).

The Clifford unitaries correspond by their action by conjugation on $\mathcal{P}_n/\langle \pm 1, \pm i\rangle$ precisely to the linear isomorphisms $\mathbb{F}_2^{2n} \to \mathbb{F}_2^{2n}$ which carry isotropic subspaces to isotropic subspaces (such a mapping is called a *symplectomorphism*). Because Clifford circuits are defined by their action on Paulis, we get the following from Theorem 3.

*Proposition 4:* Given $\xi \in \mathbb{F}_2^r$ and an ordered basis $\mathbf{x} = (x_1,\ldots,x_r)$ for an isotropic subspace of $\mathbb{F}_2^{2n}$, we can, in time poly($n$), construct an $n$-qubit Clifford $C$ with $|C| = O(n^2)$ and $C\mathsf{Z}_jC^\dagger = (-1)^{\xi_j}\mathsf{P}^{x_j}$ for each $1 \le j \le r$.

Likewise, the fact that Cliffords preserve the bilinear form also gives us the following useful observation:

*Observation 2:* Let $(x_0, y_0)$ and $(x_1, y_1)$ be pairs in $\mathbb{F}_2^{2n}$ where, for each $k = 0, 1$, $\mathsf{P}^{x_k}$ and $\mathsf{P}^{y_k}$ anticommute. Then there is an $n$-qubit Clifford $F$ such that $F\mathsf{P}^{x_0}F^\dagger = \mathsf{P}^{x_1}$ and $F\mathsf{P}^{y_0}F^\dagger = \mathsf{P}^{y_1}$.

## VI. Reductions to and From ENIC

We first note the following, which we use repeatedly and just follows from Theorem 3, by an appropriate pair of Clifford transformations.

*Observation 3:* Given $x, y, x', y' \in \mathbb{F}_2^{2n} \setminus \{\mathbf{0}\}$ and a circuit $A \in \mathcal{T}_d^n$, we can efficiently construct a circuit $B \in \mathcal{T}_d^n$, with $|B| = |A| + O(n^2)$ gates, such that $\langle A\mathsf{P}^xA^\dagger, \mathsf{P}^y\rangle = \langle B\mathsf{P}^{x'}B^\dagger, \mathsf{P}^{y'}\rangle$.

Our sequence of reductions for this section is contained in Propositions 5 and 6.

*Proposition 5:* Given as input an $n$-qubit $C \in \mathcal{T}^d$ and a nonzero $z \in \mathbb{F}_2^{2n}$, we can efficiently construct an $O(n)$-qubit Clifford+$\mathsf{T}$ circuit $F$ where:

i) $F$ has $\mathsf{T}$-depth $d + O(\log(n))$ and has $|C| + O\big(\mathrm{poly}(n)\big)$ gates; AND
ii) $\mathsf{P}^z$ appears with nonzero coefficient in the Pauli expansion of $C$ if and only if $F$ exactly instantiates identity up to global phase.

*Proof:* Let $\sum_{v\in\mathbb{F}_2^{2n}} \alpha_v \mathsf{P}^v$ be the Pauli expansion of the unitary instantiated by $C$. By Observation 3, we may suppose that $z = (\mathbf{0}_n, \mathbf{1}_n)$, i.e. $\mathsf{P}^z = \mathsf{X}^{\otimes n}$. Now we add $2n$ extra registers and then add, by conjugation, the following rounds of Clifford gates to $C \otimes \mathsf{I}^{\otimes 2n}$. This is illustrated in Figure 2.

1) First, we add the pairwise-commuting gates $\mathsf{CZ}_{n,2n}, \mathsf{CZ}_{n-1,2n-1}, \ldots, \mathsf{CZ}_{2,n+2}, \mathsf{CZ}_{1,n+1}$.
2) Then we add a round of $\mathsf{HSH}$ (i.e. $\sqrt{\mathsf{X}}$-gates) on wires $1,\ldots,n$.
3) Then we add the pairwise-commuting gates $\mathsf{CZ}_{n,3n}, \mathsf{CZ}_{n-1,3n-1}, \ldots, \mathsf{CZ}_{2,2n+2}, \mathsf{CZ}_{1,2n+1}$.
4) Finally, we add a round of Hadamards on wires $n+1, n+2, \ldots, 3n$.

The effect of the conjugation of $C \otimes \mathsf{I}^{\otimes 2n}$ in Figure 2 is that we obtain a new $3n$-qubit circuit $D$ whose corresponding unitary can be expressed in the form

$$D = \sum_{v\in\mathbb{F}_2^{2n}} \alpha_v (-1)^{f(v)}\mathsf{P}^{K(v)} \otimes \mathsf{P}^{L(v)}$$

where

1) $K : \mathbb{F}_2^{2n} \to \mathbb{F}_2^{2n}$ and $f : \mathbb{F}_2^{2n} \to \mathbb{F}_2$ are maps where $\mathsf{P}^v \to (-1)^{f(v)}\mathsf{P}^{K(v)}$ has the effect, on each of wires $1,\cdots,n$, of fixing $\mathsf{X}$ while interchanging $\mathsf{Z}$ with $-\mathsf{Y}$; AND
2) $L : \mathbb{F}_2^{2n} \to \mathbb{F}_2^{4n}$ is a map whose images lies entirely within the $\mathsf{X}$-subspace of $\mathbb{F}_2^{4n}$, and we have $\mathsf{P}^{L(v)} = \mathsf{X}^{\otimes 2n}$ if and only if $v = z$.

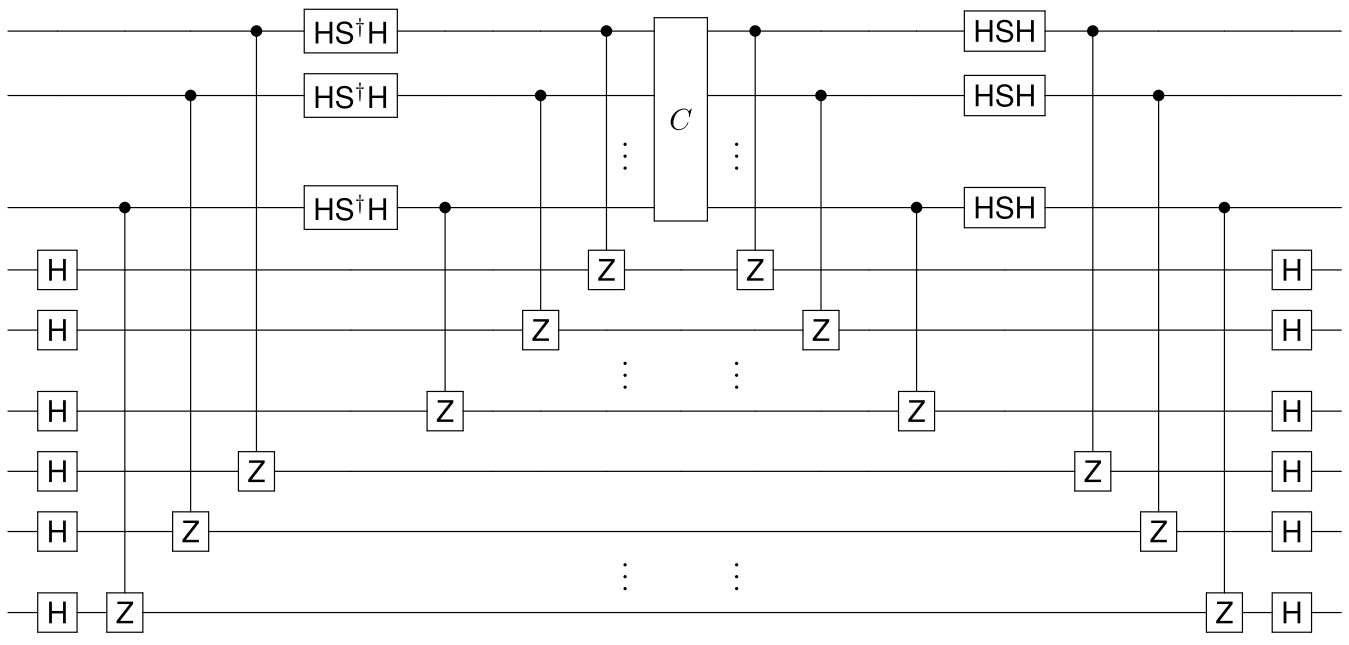

Fig. 2. Entangling $C$ with $2n$ auxiliary registers.

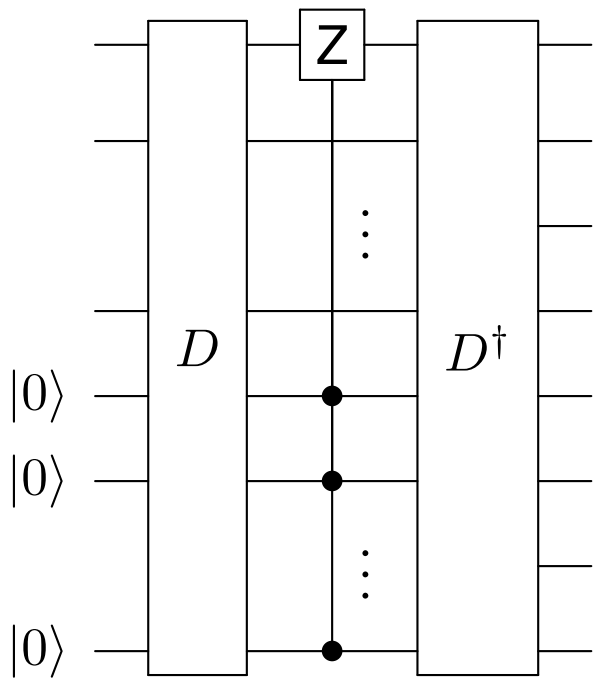

Fig. 3. The circuit $V$ detects the presence of $\mathsf{P}^z$ in the Pauli expansion of $C$.

Now we add a multi-qubit controlled-$\mathsf{Z}$ gate $M$ with control qubits $n+1,\dots,3n$ and target qubit 1 (equivalent to a $(2n+1)$-Toffoli gate up to Hadamard basis change on the target register), acting as identity on the remaining wires, and we consider the circuit $V := D^\dagger M D$ with registers $n+1, n+2,\dots,3n$ treated as ancillas all initialized to $|0\rangle$, as illustrated in Figure 3. Regarding $V$ as a circuit with $n$ data registers and $2n$ ancilla registers, we get that $V$ instantiates the identity up to global phase if and only if $\alpha_z = 0$. In particular, for any computational basis state $|b_1 \dots b_n\rangle$, we have

$$
\begin{aligned}
&V|b_1 \dots b_n\rangle|0\rangle^{\otimes 2n} = \\
&|b_1 \dots b_n\rangle|0\rangle^{\otimes 2n} - 2b_1 \cdot \alpha_z D^\dagger|(b_1 \oplus 1)\dots(b_n \oplus 1)\rangle|1\rangle^{\otimes 2n}
\end{aligned}
$$

Now, using only a single ancilla, using [19], the gate $M$ can be instantiated (exactly) as a Clifford+$\mathsf{T}$-circuit with $\mathsf{T}$-depth $O(\log(n))$ (see [20] and [21] for some additional recent constructions for optimizing the circuit cost of implementing $n$-Toffolis. In particular, we cannot do better than a $\mathsf{T}$-depth of $O(\log(n))$).

The above construction involves the use of $O(n)$ ancillas, but we can just add $O(n)$ extra registers and uncompute the garbage, tracing out the reset ancillary qubits to produce a reversible quantum circuit. This only has the effect of doubling the size of the circuit and, in particular, doubling the $\mathsf{T}$-depth, and produces a classical description of an ancilla-free $O(n)$-qubit quantum circuit which instantiates identity up to global phase if and only if $\alpha_z = 0$. □

For any instance $(C, x) \in \mathcal{T}_n^d \times \mathbb{F}_2^{2n}$ of SUPPORT, the circuit $C\mathsf{P}^x C^\dagger$ has $\mathsf{T}$-depth $\leq 2d$, so Proposition 5 implies that ENIC over domain $\mathcal{T}^{2d+O(\log(n))}$ is at least as hard as SUPPORT over domain $\mathcal{T}^d$.

*Proposition 6:*

1) For any class $\mathcal{F}$ of reversible quantum circuits, there is a polynomial-time reduction from ENIC over domain $\mathcal{F}$ to COMMUTE over domain $\mathcal{F}$; AND
For any $d > 0$, both of the following hold.
2) If there is an efficient function-determined conjugate-encoding $\mathcal{A}$ over $\mathcal{T}^d$, then COMMUTE over domain $\mathcal{T}^{2d}$ lies in P.
3) There is a polynomial-time reduction from COMMUTE over domain $\mathcal{T}^d$ to ENIC over domain $\mathcal{T}^{2d}$.

*Proof:* We first prove a. Let $D$ be an $n$-qubit circuit of $\mathcal{F}$. Since $n$-qubit Clifford unitaries are, up to global phase, uniquely determined by their action on $\mathcal{P}_n$, we get that $D \equiv \mathsf{I}^{\otimes n}$ if and only if $D\mathsf{P}^x D^\dagger = \mathsf{P}^x$ for each $x \in \mathbb{F}_2^{2n}$. If there exists an $e \in \{e_{\mathsf{Z}}^1, \dots, e_{\mathsf{Z}}^n, e_{\mathsf{X}}^1, \dots, e_{\mathsf{X}}^n\}$ such that $D\mathsf{P}^e D^\dagger \neq \mathsf{P}^e$, then $D \not\equiv \mathsf{I}^{\otimes n}$. Conversely, if no such $e$ exists, then $D\mathsf{P}^x D^\dagger = \mathsf{P}^x$ for each $x \in \mathbb{F}_2^{2n}$, so checking whether $D \equiv \mathsf{I}^{\otimes n}$ only requires us to call an oracle for COMMUTE on the $2n$ inputs $(D, e_{\mathsf{Z}}^1), \dots, (D, e_{\mathsf{X}}^n)$. This proves 1). For b, we just note that given any $(C, x) \in \mathcal{T}_n^{2d} \times \mathbb{F}_2^{2n}$, it is easy to see that we can (by undoing the conjugation of the LHS of $C\mathsf{P}^x C^\dagger = \mathsf{P}^x$ layer-by-layer) efficiently convert the problem of deciding whether $\langle C\mathsf{P}^x C^\dagger, \mathsf{P}^x\rangle = +1$ into an equivalent problem of comparing $(D, x)$ and $(D', x)$ where $D, D' \in \mathcal{T}_n^d$, to decide whether $D\mathsf{P}^x D^\dagger$ and $D'\mathsf{P}^x (D')^\dagger$ instantiate the exact same unitary. Now we prove c. We show that, for $(C, x) \in \mathcal{T}_n^d \times \mathbb{F}_2^{2n}$, we can decide COMMUTE for $(C, x)$ using two calls to an ENIC-oracle over domain $\mathcal{T}_n^{2d}$ on circuits with $O(|C| + n^2)$ gates. As in Proposition 5, we may suppose without loss of generality that $x = e_{\mathsf{Z}}^1$, i.e. $\mathsf{P}^x = \mathsf{Z}_1$. Let $U := C\mathsf{Z}_1 C^\dagger \mathsf{Z}_1$. Since $C\mathsf{Z}_1 C^\dagger$ is a unitary of order $\leq 2$ and each basis Pauli is Hermitian, every coefficient in the Pauli expansion of $C\mathsf{Z}_1 C^\dagger$ is real. Thus, $U$ is a no-instance of ENIC if and only if $U \equiv \mathsf{I}^{\otimes n}$ with global phase $\pm 1$. If $U$ is a yes-instance of ENIC, then $U \neq \mathsf{I}^{\otimes n}$, and so $\langle C\mathsf{Z}_1 C^\dagger, \mathsf{Z}_1\rangle \neq 1$ and we are done, so let $k \in \{0, 1\}$ with $U = (-1)^k \mathsf{I}^{\otimes n}$. Now, consider the circuit $V := C\mathsf{S}_1 C^\dagger \mathsf{S}_1 \in \mathcal{T}_n^{2d}$. We then have

$$
\begin{aligned}
V &= C\left(\tfrac{e^{i\pi/4}\mathsf{I}^{\otimes n} + e^{-i\pi/4}\mathsf{Z}_1}{\sqrt{2}}\right) C^\dagger \mathsf{S}_1 \\
&= \left(\tfrac{e^{i\pi/4}\mathsf{I}^{\otimes n} + (-1)^k e^{-i\pi/4}\mathsf{Z}_1}{\sqrt{2}}\right) \mathsf{S}_1
\end{aligned}
$$

Thus, if $k = 1$, then $V = \mathsf{I}^{\otimes n}$, and if $k = 0$, then $V = \mathsf{Z}_1$, so we can distinguish these cases with one call of an ENIC-oracle to $V$, and we are done. This proves c. □

## VII. Encoding Linear Combinations of Pauli Coefficients of PC-Circuits

This section consists of three lemmas in which we show how to perform arithmetic operations using Pauli coefficients of low-$\mathsf{T}$-depth PC circuits, where the latter two lemmas are essentially immediate consequences of the first. The main task we face is as follows: Suppose we are given two $n$-qubit circuits $C, D$, each with a $\mathsf{T}$-depth $\leq d$, with $\langle C\mathsf{Z}_1 C^\dagger, \mathsf{Z}_1\rangle = \alpha$ and $\langle D\mathsf{Z}_1 D^\dagger, \mathsf{Z}_1\rangle = \beta$. How do we use $C, D$ to construct a circuit $U$ such that the PC-circuit $U\mathsf{Z}_1 U^\dagger$ has $\mathsf{Z}_1$-coefficient $\frac{\alpha+\beta}{2}$, where $U$ has $\mathsf{T}$-depth at most $d + O(1)$? In particular, we are not permitted to compose $C$ and $D$, although we can take their tensor product. Our strategy is as follows:

i) We use a single extra $\mathsf{T}$-gate conjugation, together with some Cliffords and extra registers, to create an intermediate circuit which has a coefficient of $\alpha/\sqrt{2}$ on one Pauli and a coefficient of $-\beta/\sqrt{2}$ on a different Pauli, where these two Paulis anticommute.
ii) We shift that circuit by another Clifford conjugation so as to produce a circuit whose $\mathsf{X}_1$-coefficient is $\alpha/\sqrt{2}$ and whose $\mathsf{Y}_1$-coefficient is $-\beta/\sqrt{2}$.
iii) Now one more $\mathsf{T}_1$-gate conjugation has the effect of averaging out these two branches to produce a circuit whose $\mathsf{X}_1$-coefficient is $\frac{\alpha-\beta}{2}$ and whose $\mathsf{Y}_1$-coefficient is $-\frac{\alpha+\beta}{2}$. We can then correct this with one final Clifford.

We make this precise below.

*Lemma 1:* Let $C, D \in \mathcal{T}_n^d$ and $x \in \mathbb{F}_2^{2n} \setminus \{\mathbf{0}\}$. Set $\alpha := \left\langle C\mathsf{P}^x C^\dagger, \mathsf{P}^x\right\rangle$ and $\beta := \left\langle D\mathsf{P}^x D^\dagger, \mathsf{P}^x\right\rangle$. Then both of the following hold.

1) We can efficiently construct a $2n$-qubit circuit $U' \in \mathcal{T}^d$ with $|U'| = O(|C| + |D|)$ and $\langle U'\mathsf{Z}_1(U')^\dagger, \mathsf{Z}_1\rangle = \alpha\beta$.
2) We can efficiently construct an $(2n+1)$-qubit circuit $U \in \mathcal{T}^{d+2}$, with $|U| = O(|C| + |D|)$ and $\langle U\mathsf{Z}_1 U^\dagger, \mathsf{Z}_1\rangle = \frac{\alpha+\beta}{2}$.

*Proof:* 1) is straightforward. We just apply Observation 3 to a tensor product of $C$ and $D$. Now we prove 2). We construct $U$ in stages. As in Proposition 5, we may suppose that $x = e^1_{\mathsf{Z}}$. Firstly, there exist $2^{n-1} \times 2^{n-1}$ complex matrices $M_1, M_2, M_3, M_4$ such that

$$C\mathsf{P}^x C^\dagger = \mathsf{Z} \otimes M_1 + \mathsf{X} \otimes M_2 + \mathsf{Y} \otimes M_3 + \mathsf{I} \otimes M_4$$

where, in particular, $\left\langle M_1, \mathsf{I}^{\otimes n-1}\right\rangle = \alpha$. To continue, we now show that we can construct an $(n+1)$-qubit circuit $Q \in \mathcal{T}^d$ such that $Q$ commutes with $\mathsf{Y}_1$ and $\left\langle Q\mathsf{X}_1 Q^\dagger, \mathsf{Z}_1\right\rangle = \alpha$. We first note that, by Observation 2, we can find an $(n+1)$-qubit Clifford $F$, acting only one wires $1, n+1$, such that $F\mathsf{X}_1 F^\dagger = \mathsf{Z}_1\mathsf{X}_{n+1}$ and $F\mathsf{Y}_1 F^\dagger = \mathsf{Z}_{n+1}$. Likewise, we can find an $(n+1)$-qubit Clifford $G$, acting only on wires $1, n+1$, such that $G\mathsf{Z}_1\mathsf{X}_{n+1}G^\dagger = \mathsf{Z}_1$ and $G\mathsf{Z}_{n+1}G^\dagger = \mathsf{Y}_1$. We now set $Q := G(C \otimes \mathsf{I})F$. The action by conjugation by $Q$ gives us $Q\mathsf{Y}_1 Q^\dagger = \mathsf{Y}_1$, so $Q$ commutes with $\mathsf{Y}_1$. Now, we have

$$(C \otimes \mathsf{I})\mathsf{Z}_1(C \otimes \mathsf{I})^\dagger = \mathsf{Z} \otimes M_1 \otimes \mathsf{I} + \mathsf{X} \otimes M_2 \otimes \mathsf{I} + \mathsf{Y} \otimes M_3 \otimes \mathsf{I} + \mathsf{I} \otimes M_4 \otimes \mathsf{I}$$

and $(C \otimes \mathsf{I})F\mathsf{X}_1 F^\dagger(C \otimes \mathsf{I})^\dagger = (C\mathsf{P}^x C^\dagger) \otimes \mathsf{X}$, so we get

$$(C \otimes \mathsf{I})F\mathsf{X}_1 F^\dagger(C \otimes \mathsf{I})^\dagger = \mathsf{Z} \otimes M_1 \otimes \mathsf{X} + \mathsf{X} \otimes M_2 \otimes \mathsf{X} + \mathsf{Y} \otimes M_3 \otimes \mathsf{X} + \mathsf{I} \otimes M_4 \otimes \mathsf{X}$$

So, conjugating by $G$, we indeed have $\left\langle Q\mathsf{X}_1 Q^\dagger, \mathsf{Z}_1\right\rangle = \alpha$. With this $Q$ in hand, we now define the following.

1) Again applying Observation 2, we fix a $(2n+1)$-qubit Clifford $A$, acting only on wires $1, n+2$, such that both $A\mathsf{X}_1 A^\dagger = \mathsf{X}_1$ and $A\mathsf{Y}_1 A^\dagger = \mathsf{Y}_1\mathsf{Z}_{n+2}$.
2) Likewise, we fix a $(2n+1)$-qubit Clifford $B$, acting only on wires $1, n+2$, such that $B\mathsf{Z}_1 B^\dagger = \mathsf{X}_1$ and $B\mathsf{Y}_1\mathsf{Z}_{n+2}B^\dagger = \mathsf{Y}_1$.
3) Let $V$ denote the circuit $Q \otimes D$. Note that this has $\mathsf{T}$-depth $\leq d$.
4) Now we set $U := \mathsf{H}_1\mathsf{S}_1\mathsf{T}_1 BVA\mathsf{T}_1\mathsf{H}_1 \in \mathcal{T}^{d+2}_{2n+1}$.

We check in stages that $\left\langle U\mathsf{Z}_1 U^\dagger, \mathsf{Z}_1\right\rangle = \frac{\alpha+\beta}{2}$. Recall that $\mathsf{THZHT}^\dagger = \frac{\mathsf{X}-\mathsf{Y}}{\sqrt{2}}$. Since $Q$ commutes with $\mathsf{Y}_1$ and does not act on wire $n+2$, we have

$$\frac{V\mathsf{X}_1 V^\dagger - V\mathsf{Y}_1\mathsf{Z}_{n+2}V^\dagger}{\sqrt{2}} = \frac{(Q \otimes \mathsf{I}^{\otimes n})\mathsf{X}_1(Q \otimes \mathsf{I}^{\otimes n})^\dagger - \mathsf{Y} \otimes \mathsf{I}^{\otimes n} \otimes (D\mathsf{Z}_1 D^\dagger)}{\sqrt{2}} \tag{3}$$

and we have

$$\left\langle V\mathsf{X}_1 V^\dagger, \mathsf{Z}_1\right\rangle = \alpha \text{ and } \left\langle V\mathsf{X}_1 V^\dagger, \mathsf{Y}_1\mathsf{Z}_{n+2}\right\rangle = 0$$

$$\left\langle V\mathsf{Y}_1\mathsf{Z}_{n+2}V^\dagger, \mathsf{Y}_1\mathsf{Z}_{n+2}\right\rangle = \beta \text{ and } \left\langle V\mathsf{Y}_1\mathsf{Z}_{n+2}V^\dagger, \mathsf{Z}_1\right\rangle = 0$$

So conjugating by $B$ gives us

$$\left\langle BV\mathsf{X}_1 V^\dagger B^\dagger, \mathsf{X}_1\right\rangle = \alpha \text{ and } \left\langle BV\mathsf{X}_1 V^\dagger B^\dagger, \mathsf{Y}_1\right\rangle = 0$$

$$\left\langle BV\mathsf{Y}_1\mathsf{Z}_{n+2}V^\dagger B^\dagger, \mathsf{Y}_1\right\rangle = \beta \text{ and } \left\langle BV\mathsf{Y}_1\mathsf{Z}_{n+2}V^\dagger B^\dagger, \mathsf{X}_1\right\rangle = 0$$

Now we apply the next $\mathsf{T}_1$. Since $\mathsf{T}$ commutes with $\mathsf{Z}$ and, in particular, $\mathsf{TYT}^\dagger = i\mathsf{TXT}^\dagger\mathsf{Z} = i\left(\frac{\mathsf{X}-\mathsf{Y}}{\sqrt{2}}\right)\mathsf{Z} = \frac{\mathsf{Y}+\mathsf{X}}{\sqrt{2}}$, we get

$$\left\langle \mathsf{T}_1 BV\mathsf{X}_1 V^\dagger B^\dagger\mathsf{T}_1^\dagger, \mathsf{X}_1\right\rangle = \alpha/\sqrt{2}$$
$$\left\langle \mathsf{T}_1 BV\mathsf{X}_1 V^\dagger B^\dagger\mathsf{T}_1^\dagger, \mathsf{Y}_1\right\rangle = -\alpha/\sqrt{2}$$
$$\left\langle \mathsf{T}_1 BV\mathsf{Y}_1\mathsf{Z}_{n+2}V^\dagger B^\dagger\mathsf{T}_1^\dagger, \mathsf{X}_1\right\rangle = \beta/\sqrt{2}$$
$$\left\langle \mathsf{T}_1 BV\mathsf{Y}_1\mathsf{Z}_{n+2}V^\dagger B^\dagger\mathsf{T}_1^\dagger, \mathsf{Y}_1\right\rangle = \beta/\sqrt{2}$$

So, our application of $\mathsf{H}_1\mathsf{S}_1\mathsf{T}_1 B$ by conjugation to (3) gives us $\langle U\mathsf{Z}_1 U^\dagger, \mathsf{Z}_1\rangle = \frac{\alpha+\beta}{2}$, and we are done. □

*Lemma 2:* Let $x \in \mathbb{F}_2^{2n} \setminus \{\mathbf{0}\}$ and $C_1, \cdots, C_k$ be a family of $n$-qubit circuits, all of $\mathsf{T}$-depth $\leq d$. Then we can efficiently construct an $O(kn)$-qubit Clifford+$\mathsf{T}$ circuit $U$ of $\mathsf{T}$-depth $\leq d + 2\log(k)$, such that $|U| = O(|C_1| + \cdots + |C_k|)$ and

$$\langle U\mathsf{Z}_1 U^\dagger, \mathsf{Z}_1\rangle = \frac{1}{2^{\lceil \log_2(k)\rceil}} \sum_{s=1}^{k} \langle C_s\mathsf{P}^x C_s^\dagger, \mathsf{P}^x\rangle$$

*Proof:* It suffices to prove that this holds in the case where $k$ is a power of two because, for an arbitrary $k$ we can pad the list $C_1, \ldots, C_k$ with at most $k$ extra Clifford circuits $C$, where $\langle C\mathsf{P}^x C^\dagger, \mathsf{P}^x\rangle = 0$, until this holds. In the case where $k$ is a power of two, it is clear that Lemma 2 just follows from 2) of Lemma 1 above. In particular, each time we double the number of terms in the linear combination, we increase the $\mathsf{T}$-depth by two and enlarge the number of registers by a constant factor. □

We now define the following.

*Definition 19:* Given an $r \in \mathbb{Z}[\sqrt{2}]$, we define a *generalized base-$\sqrt{2}$ expression* for $r$ to be a pair $\left(\vec{a}, \vec{b}\right)$ of finite binary strings of the same length $k$, such that

$$r = \sum_{j=0}^{k} (-1)^{a_j} b_j (\sqrt{2})^j$$

Note that such an expression for $r$ is not necessarily unique. We call $k$ the *length* of the generalized base-$\sqrt{2}$ expression. Now, Lemma 1 immediately yields the following.

*Lemma 3:*

1) Given as input an integer $n \geq k \geq 1$, we can efficiently construct an $n$-qubit Clifford PC-circuit of size $|C| = O(k)$, where $\langle C, \mathsf{Z}_1 \rangle = \pm 2^{-k/2}$.
2) Given as input a $d \in \mathbb{Z}[\sqrt{2}]$ presented as a length-$m$ generalized base-$\sqrt{2}$-expression, we can efficiently construct an $O(m)$-qubit PC-circuit $C$ with $\mathsf{T}$-depth $O(\log(m))$ and size $|C| = O(m^2)$, where $\langle C, \mathsf{Z}_1 \rangle = \frac{d}{2^\ell}$ and $\ell = O(m^2)$.

*Proof:* 1) is immediate, since, using Observation 3, we can just take our initial Pauli to be $\mathsf{P}^x = \mathsf{X}^{\otimes k}$, conjugate by a single layer of $\mathsf{T}$-gates, and apply another appropriate Pauli conjugation to get a sign flip, recalling that conjugates of Paulis by a single layer of $\mathsf{T}$-gates are Clifford. For 2), we just combine 1) with Lemma 2, and we are done. □

## VIII. Depth-$d$ Presentations

For each $d \geq 0$, we now define a set $\mathcal{E}_d$ of simple data structures called *depth-d presentations*. Each element $\Lambda$ of $\mathcal{E}_d$ can be regarded as a binary string which encodes precisely one unitary $U^\Lambda$, where $U$ admits a gate description of the form $C\mathsf{P}^x C^\dagger$ for $(C, x) \in \mathcal{T}_n^d \times \mathbb{F}_2^{2n}$. Conversely, each circuit of the form $C\mathsf{P}^x C^\dagger$ for $(C, x) \in \mathcal{T}_n^d \times \mathbb{F}_2^{2n}$ can be converted into a depth-$d$ presentation $\Lambda$ with $U^\Lambda = C\mathsf{P}^x C^\dagger$. We will see in Section IX that these data structures are a natural way to write down all the information we need to describe the coefficients of the Pauli expansions of PC-circuits.

*Definition 20:* Given an $m \geq 1$ and an ordered basis $\mathbf{x}$ for a subspace $W \subseteq \mathbb{F}_2^m$, we write $\dim(\mathbf{x})$ to denote $\dim(W)$ and we let $W_\mathbf{x}$ denote the subspace of $\mathbb{F}_2^m$ generated by $\mathbf{x}$. If $\mathbf{x}$ is an empty basis, then $W_\mathbf{x} = \{\mathbf{0}\}$. For $d \geq 0$, a *depth-d presentation* is a tuple $\Lambda = ((\mathbf{x}_d, \sigma_d), (\mathbf{x}_{d-1}, \sigma_{d-1}), \ldots, (\mathbf{x}_1, \sigma_1), (y, \tau))$ where

1) $(y, \tau) \in \mathbb{F}_2^{2n} \times \mathbb{F}_2 \setminus \{(\mathbf{0}_{2n}, 1)\}$; AND
2) For each $1 \leq j \leq d$, $\mathbf{x}_j$ is an ordered basis for an isotropic subspace of $\mathbb{F}_2^{2n}$ and $\sigma_j \in \mathbb{F}_2^{\dim(\mathbf{x}_j)}$.

We call $y$ the *outer vector* of $\Lambda$ and $(y, \tau)$ the *phased outer vector* of $\Lambda$. Forbidding $(\mathbf{0}_{2n}, 1)$ just ensures that $\Lambda$ is never an encoding of the unitary $-\mathsf{I}^{\otimes n}$, which does not admit a PC-circuit gate description. We usually think of each $\sigma_j \in \mathbb{F}_2^{\dim(\mathbf{x}_j)}$ as a linear map from $W_{\mathbf{x}_j} \to \mathbb{F}_2$, defined by its action on the basis vectors. For $v \in W_{\mathbf{x}_j}$, we write $\sigma_j(v)$ accordingly. We say that $\Lambda$ is a *presentation* if it is an element of $\bigcup_{d\geq 0} \mathcal{E}_d$. If we want to make $n$ explicit we call $\Lambda$ an *n-qubit presentation*.

*Definition 21:* Let $A$ be an $n$-qubit Clifford circuit, $\mathbf{x}$ be an ordered basis of an isotropic subspace of $\mathbb{F}_2^{2n}$, and $\sigma \in \mathbb{F}_2^{\dim(\mathbf{x})}$. We write $A(\mathbf{x}, \sigma)$ to denote the pair $(\mathbf{u}, \xi)$ obtained by updating $(\mathbf{x}, \sigma)$ by conjugation by $A$. That is, letting $\mathbf{x} = (x_1, \ldots, x_r)$, we have $\mathbf{u} = (u_1, \ldots, u_r)$ and $\xi \in \mathbb{F}_2^r$, where $(-1)^{\xi(u_j)}\mathsf{P}^{u_j} = (-1)^{\sigma(x_j)} A \mathsf{P}^{x_j} A^\dagger$ for each $1 \leq j \leq r$.

There is a simple algorithm which, given an input $(C, x) \in \mathcal{T}_n^d \times \mathbb{F}_2^{2n}$, outputs a depth-$d$ presentation.

*Proposition 7:* There is an algorithm $\mathcal{A}^{\text{enc}} : \mathcal{T}_n \times \mathbb{F}_2^{2n} \to \bigcup_{d\geq 0} \mathcal{E}_d$, where, given $(C, x) \in \mathcal{T}_n^d \times \mathbb{F}_2^{2n}$ as input, $\mathcal{A}^{\text{enc}}$ outputs a $\Lambda \in \mathcal{E}_d$, where $\mathcal{A}^{\text{enc}}$ runs in time $O((d+1)^2 n^2 |C|)$. Furthermore, $\mathcal{A}^{\text{enc}}(C, x)$ has the all-zero vector as its outer vector if and only if $x = \mathbf{0}$.

*Proof:* We write the circuit $C$ as $C = F_d T_d F_{d-1} T_{d-1} \cdots F_1 T_1 F_0$, where $F_0, \cdots, F_d$ are Cliffords and, for each $j = 1, \ldots, d$, there is a vector $(s_{j1}, \ldots, s_{jn}) \in \mathbb{F}_2^n$ such that $T_j = \mathsf{T}^{s_{j1}} \otimes \cdots \otimes \mathsf{T}^{s_{jd}}$. Let $\mathbf{T}_j$ be the ordered subbasis $(e_\mathsf{Z}^r : s_{jr} = 1)$ of $\mathbf{E}_\mathsf{Z}^n$ and $h_j$ be the length of this subbasis. Then $\Lambda$ is obtained in the following way:

1) We begin with the depth-zero presentation $\Lambda_0 = ((x, 0))$.
2) Now we apply $F_0$ to $(x, 0)$ by conjugation (as in Definition 21) to get a new depth-zero presentation $\Lambda_0' = ((x', \lambda'))$.
3) Then we add $(\mathbf{T}_1, \mathbf{0}_{h_1})$ to the left of $\Lambda_0'$ to get depth-one presentation $\Lambda_1 = ((\mathbf{T}_1, \mathbf{0}_{h_1}), (x', \lambda'))$.
4) Then we conjugate each of $(\mathbf{T}_1, \mathbf{0}_{h_1})$ and $(x', \lambda')$ by $F_1$ to get a new depth-one presentation $\Lambda_1'$.
5) Then we add $(\mathbf{T}_2, \mathbf{0}_{h_2})$ to the left of $\Lambda_1'$ to get depth-two presentation $\Lambda_2$.
6) Then we conjugate each pair in $\Lambda_2$ by $F_2$ as in 4) to get a new depth-two presentation $\Lambda_2'$.
7) We proceed in this way until we have passed through the entire circuit $C$ and have a final depth-$d$ presentation $\Lambda = \Lambda_d'$. □

We now show that the procedure above admits a decoding.

*Proposition 8:* There is a procedure $\mathcal{R}$ such that

1) $\mathcal{R}$ takes as input pairs $(\Lambda, z) \in \left(\bigcup_{d\geq 0} \mathcal{E}_d\right) \times \mathbb{F}_2^{2n}$ such that, letting $\Lambda$ have outer vector $y$, we have $y = \mathbf{0}$ if and only if $z = \mathbf{0}$; AND
2) Given a $\Lambda \in \mathcal{E}_d$, the following hold:
   a) On input $(\Lambda, z)$, the output of $\mathcal{R}$ is a circuit $C \in \mathcal{T}_n^d$ such that $\mathcal{A}^{\text{enc}}(C, z) = \Lambda$ and $|C| = O(n^2(d+1))$, and $\mathcal{R}$ runs in time $(d+1)^2 \cdot \text{poly}(n)$.
   b) There is a unitary $U^\Lambda$, depending only on $\Lambda$, such that $U^\Lambda = (\mathcal{R}(\Lambda, z))\mathsf{P}^z(\mathcal{R}(\Lambda, z))^\dagger$ for any $z \in \mathbb{F}_2^{2n}$ satisfying 1). In particular, if $z = \mathbf{0}$, then $U^\Lambda = \mathsf{I}^{\otimes n}$.

*Proof:* We show this recursively, simply running the procedure of Proposition 7 in reverse. Given an input pair $(\Lambda, z)$, where $\Lambda$ is a depth-zero presentation $(y, \tau)$, we have a trivial algorithm. Applying Proposition 4, we find a Clifford circuit $A$ such that $A\mathsf{P}^z A^\dagger = (-1)^\tau \mathsf{P}^y$ and we output $A$. Now let $d \geq 1$ suppose that $\mathcal{R}$ is already defined over input pairs $(\Lambda, z)$ where $\Lambda$ has depth $\leq d-1$, and that $\mathcal{R}$ satisfies the specified conditions over these input pairs. Let $\Lambda := ((\mathbf{x}_d, \sigma_d), \cdots, (\mathbf{x}_1, \sigma_1), (y, \tau))$. Let $\dim(\mathbf{x}_d) = r$, where $\mathbf{x}_d = (x_{d1}, \ldots, x_{dr})$. Applying Proposition 4 again, we find a Clifford circuit $A$ such that $A^\dagger(\mathbf{x}_d, \sigma_d) = (\mathbf{E}_\mathsf{Z}^r, \mathbf{0})$. Let $\Lambda'$ be the depth-$(d-1)$ presentation $(A^\dagger(\mathbf{x}_{d-1}, \sigma_{d-1}), \ldots, A^\dagger(\mathbf{x}_1, \sigma_1), A^\dagger(y, \tau))$. We define $U^\Lambda$ to be the unitary $AT(U^{\Lambda'})T^\dagger A^\dagger$. Furthermore, given a $z \in \mathbb{F}_2^{2n}$ satisfying 1), and a $C \in \mathcal{T}_n^{d-1}$, where $C\mathsf{P}^z C^\dagger = \mathcal{R}(\Lambda', z)$, we then define $\mathcal{R}(\Lambda, z)$ to be the circuit $ATC$. With these definitions, $\mathcal{R}$ still satisfies 2). Furthermore, by definition, we have $U^\Lambda = (\mathcal{R}(\Lambda, z))\mathsf{P}^z(\mathcal{R}(\Lambda, z))^\dagger$. Lastly, if $z = \mathbf{0}$, then $U^{\Lambda'} = \mathsf{I}^{\otimes n}$ so $U^\Lambda = \mathsf{I}^{\otimes n}$. □

The procedure in Proposition 8 gives us the following recursive circuit-independent description of $U^\Lambda$, which is straightforward to check by induction, using the identity in Observation 1. In particular, Proposition 9 is just a generalization of Equation 2. Recall that the notation $B(\cdot,\cdot)$ always refers to the bilinear form from Section V.

*Proposition 9:* Given a depth-zero presentation $\Lambda = (y_0, \tau)$, we have $U^\Lambda = (-1)^\tau \mathsf{P}^{y_0}$. For $d \geq 1$, given a depth-$d$ presentation $\Lambda = ((\mathbf{x}_d, \sigma_d), (\mathbf{x}_{d-1}, \sigma_{d-1}), \ldots, (\mathbf{x}_1, \sigma_1), (y_0, \tau))$, we define the following:

1) Let $\Lambda_*$ be the depth-$(d-1)$ presentation $((\mathbf{x}_d, \sigma_d), \ldots, (\mathbf{x}_2, \sigma_2), (y_0, \tau))$.
2) For each basis vector $v$ in $\mathbf{x}_1$, let $\Delta_v$ be the depth-$(d-1)$ presentation $((\mathbf{x}_d, \sigma_d), \cdots, (\mathbf{x}_2, \sigma_2), (v, \sigma_1(v)))$.

In particular, if $d = 1$, then $\Lambda_* = ((y_0, \tau))$ and $\Delta_v = ((v, \sigma_1(v)))$, each a depth-zero presentation. In any case, we have:

$$U^\Lambda = \left( \prod_{\substack{v \in \mathbf{x}_1 \\ B(v, y_0) \equiv 1 \bmod 2}} \left( \frac{\mathsf{I}^{\otimes n} + iU^{\Delta_v}}{\sqrt{2}} \right) \right) U^{\Lambda_*}$$

where the unitaries $\{U^{\Delta_v} : v \in \mathbf{x}_1 \text{ and } B(v, y_0) \equiv 1 \bmod 2\}$ are pairwise-commutative and they all anticommute with $U^{\Lambda_*}$. In particular, if $\mathsf{P}^v$ commutes with $\mathsf{P}^{y_0}$ for each $v \in \mathbf{x}_1$, then $U^\Lambda = U^{\Lambda^*}$.

Consider the following example:

1) Let $y = e^1_{\mathsf{Z}} \oplus e^2_{\mathsf{X}} \oplus e^n_{\mathsf{X}} = ((1, 0, \ldots, 0), (0, 1, \ldots, 1))$, so $\mathsf{P}^y = \mathsf{Z} \otimes \mathsf{X}^{\otimes n-1}$
2) Let $\mathbf{x} = \mathbf{E}^3_{\mathsf{Z}}$ and $\sigma = (0, 0, 1)$.
3) Let $u_1 = e^1_{\mathsf{Z}} \oplus e^2_{\mathsf{Z}}$ and $u_2 = e^1_{\mathsf{X}} \oplus e^2_{\mathsf{X}}$, corresponding to the commuting Paulis $\mathsf{P}^{u_1} = \mathsf{Z}_1\mathsf{Z}_2$ and $\mathsf{P}^{u_2} = \mathsf{X}_1\mathsf{X}_2$. Let $\mathbf{u} = (u_1, u_2)$ and $\sigma' = (1, 0)$.

Then, letting $\Lambda = ((\mathbf{u}, \sigma'), (\mathbf{x}, \sigma), (y, 0))$, we have the following product decomposition of $U^\Lambda$:

$$U^\Lambda = \left( \frac{1}{2} \left( \mathsf{I}^{\otimes n} + \frac{i}{\sqrt{2}} \left( \mathsf{I}^{\otimes n} + i\mathsf{X}_1\mathsf{X}_2 \right) \mathsf{Z}_2 \right) \left( \mathsf{I}^{\otimes n} - i\mathsf{Z}_3 \right) \right) \cdot \left( \frac{1}{2} \left( \mathsf{I}^{\otimes n} - i\mathsf{Z}_1\mathsf{Z}_2 \right) \left( \mathsf{I}^{\otimes n} + i\mathsf{X}_1\mathsf{X}_2 \right) \right) \left( \mathsf{Z} \otimes \mathsf{X}^{\otimes n-1} \right)$$

In general, $\Lambda$ can simply be viewed a recording of the alternating sequence of commuting and anticommuting relations between families of Paulis that arise when we conjugate, layer-by-layer, a single Pauli by a Clifford+T circuit.

## IX. The Coefficient Space of $U^\Lambda$

We need a bit more machinery in order to relate the Pauli coefficients of $U^\Lambda$ to weight enumerator polynomials of binary linear codes in Section XI. Our first task is to write an explicit formula, in terms of $\Lambda$, for the Pauli coefficients of $U^\Lambda$, which we do at the end of this section in Equation (8). This presents some subtleties even for a conjugate of a Pauli by a T-depth-one circuit. For a subspace $V$ of $\mathbb{F}_2^{2n}$, even if $V$ is isotropic, the corresponding injection $V \to \mathcal{P}_n$ defined by $v \to \mathsf{P}^v$ is not necessarily a homomorphism. For example, if we take $V \subseteq \mathbb{F}_2^4$ to be generated by $\{(1, 1, 0, 0), (0, 0, 1, 1)\}$, the corresponding 2-qubit Paulis $\mathsf{X} \otimes \mathsf{X}$ and $\mathsf{Z} \otimes \mathsf{Z}$ commute, but $\mathsf{P}^{(1,1,1,1)} = \mathsf{Y} \otimes \mathsf{Y} = -(\mathsf{X} \otimes \mathsf{X})(\mathsf{Z} \otimes \mathsf{Z})$. In general, we can keep track of the exact phase relating $\mathsf{P}^{u \oplus v}$ to $\mathsf{P}^u\mathsf{P}^v$ using the following algebraic relation.

*Proposition 10:* For each $u \in \mathbb{F}_2^{2n}$, let $S^u_{\mathsf{Z}}, S^u_{\mathsf{X}} \subseteq [n]$, where $S^u_{\mathsf{Z}} := \{j \in [n] : (u_{\mathsf{Z}})_j \equiv 1 \mod 2\}$ and $S^u_{\mathsf{X}} := \{j \in [n] : (u_{\mathsf{X}})_j \equiv 1 \mod 2\}$. Set $[\cdot,\cdot] : \mathbb{F}_2^{2n} \times \mathbb{F}_2^{2n} \to \mathbb{Z}$ to be the bilinear form $[u, v] := |S^u_{\mathsf{Z}} \cap S^v_{\mathsf{X}}| - |S^v_{\mathsf{Z}} \cap S^u_{\mathsf{X}}| = u_{\mathsf{Z}} \cdot v_{\mathsf{X}} - v_{\mathsf{Z}} \cdot u_{\mathsf{X}}$, where the dot product is read as an integer. Then $\mathsf{P}^{u \oplus v} = i^{[u,v]}\mathsf{P}^u\mathsf{P}^v$.

The symplectic representation clearly still gives rise to isomorphisms between isotropic subspaces of $\mathbb{F}_2^{2n}$ and the corresponding stabilizer subgroups of $\mathcal{P}_n$.[1] For example, with $V$ as above, we have an obvious isomorphism between $V = \{(0, 0, 0, 0), (1, 1, 0, 0), (0, 0, 1, 1), (1, 1, 1, 1)\}$ and $\{\mathsf{I}^{\otimes 2}, \mathsf{Z} \otimes \mathsf{Z}, \mathsf{X} \otimes \mathsf{X}, -\mathsf{Y} \otimes \mathsf{Y}\}$. More generally, we define the following:

*Definition 22:* Let $\mathbf{x} = (x_1, \ldots, x_r)$ be an ordered basis for an isotropic subspace $V$ of $\mathbb{F}_2^{2n}$. Then there is a map $\theta_{\mathbf{x}} : V \to \mathbb{F}_2^{2n}$ where, for $(s_1, \ldots, s_r) \in \mathbb{F}_2^r$ and $v \in V$ with $v = \bigoplus s_j x_j$, we have $\prod_{j=1}^r \mathsf{P}^{s_j x_j} = (-1)^{\theta_{\mathbf{x}}(v)}\mathsf{P}^v$.

This map is clearly not $\mathbb{F}_2$-linear in general (in particular, each of $x_1, \cdots, x_r$ maps to zero but in general, $\theta_{\mathbf{x}}(v)$ is not necessarily zero), but the following observation is useful:

*Observation 4:* With $\mathbf{x}, V$ as in Definition 22, if $V$ is either an X- or Z-subspace, then $\theta_{\mathbf{x}}$ is identically zero.

We use the map in Definition 22 to keep track of phases while we decompose $U^\Lambda$ into a linear combination of Paulis. Additionally, when we perform this decomposition, for each isotropic subspace base $\mathbf{x}$ in $\Lambda$, we need to keep track of the $\mathbb{F}_2$-coordinate representation (in the $\mathbf{x}$-basis) of each $v \in W_{\mathbf{x}}$. This is the motivation for the additional notation we introduce below.

*Definition 23:*

1) Given $m \geq 1$, an ordered basis $\mathbf{x} = (x_1, \ldots, x_k)$ of a subspace of $\mathbb{F}_2^m$, and a $v \in W_{\mathbf{x}}$, letting $\bigoplus_{j \in S} x_j$ be the unique expression of $v$ in the $\mathbf{x}$-basis, we define $\omega(\mathbf{x}, v) := |S|$, i.e. the $\mathbf{x}$-basis analogue to the Hamming weight of $v$. Furthermore, given a $z \in \mathbb{F}_2^{\dim(\mathbf{x})}$, we set $W^z_{\mathbf{x}} := \mathrm{Span}(x_j : z_j = 1)$. In particular, $W^{(0,0,\ldots,0)}_{\mathbf{x}} = \{\mathbf{0}\}$ and $W^{(1,1,\ldots,1)}_{\mathbf{x}} = W_{\mathbf{x}}$.
2) Let $\mathbf{x} = (x_1, \ldots, x_k)$ be an ordered basis for an isotropic subspace of $\mathbb{F}_2^{2n}$. We write $B_{\mathbf{x}}$ to mean the linear map $\mathbb{F}_2^{2n} \to \mathbb{F}_2^{\dim(\mathbf{x})}$ defined by

$$v \to \begin{pmatrix} B(x_1, v) \\ \vdots \\ B(x_k, v) \end{pmatrix}$$

In particular, note that $W_{\mathbf{x}} \subseteq \ker(B_{\mathbf{x}})$. For $y \in \mathbb{F}_2^{2n}$, we call $(\mathbf{x}, y)$ an *anticommutation pair* if $B_{\mathbf{x}}(y)$ is the all-1s vector, i.e. $\mathsf{P}^y$ anticommutes with $\mathsf{P}^v$ for each basis vector $v$ of $\mathbf{x}$. We write $\mathrm{Anti}(\mathbf{x})$ to mean the set of $y \in \mathbb{F}_2^{2n}$ such that $(\mathbf{x}, y)$ is an anticommutation pair. Note that $\mathrm{Anti}(\mathbf{x})$ is an affine space of dimension $2n - \dim(\mathbf{x})$.

[1] Recall that a *stabilizer subgroup* of $\mathcal{P}_n$ is an abelian subgroup $S$ whose elements are all Hermitian, where $-\mathsf{I}^{\otimes n} \notin S$.

Note that symplectomorphisms carry anticommutation pairs to anticommutation pairs. With the notation above in hand, we can describe precisely how a depth-one presentation expands into a uniform-magnitude linear combination of depth-zero presentations (i.e. Paulis).

*Proposition 11:* Let $\Lambda = ((\mathbf{x}_1, \sigma_1), (y_0, \tau))$ be a depth-one presentation. Then

$$2^{|B_{\mathbf{x}_1}(y_0)|/2} U^\Lambda = \sum_{z \in W_{\mathbf{x}_1}^{B_{\mathbf{x}_1}(y_0)}} (-1)^{\tau \oplus \sigma_1(z) \oplus \theta_{\mathbf{x}_1}(z)} i^{\omega(\mathbf{x}_1, z) - [z, y_0]} \mathsf{P}^{z \oplus y_0} \tag{4}$$

*Proof:* If all the Paulis corresponding to the basis entries of $\mathbf{x}_1$ commute with $\mathsf{P}^{y_0}$, then $W_{\mathbf{x}_1}^{B_{\mathbf{x}_1}(y_0)} = \{\mathbf{0}\}$ and $U^\Lambda = (-1)^\tau \mathsf{P}^{y_0}$, so we are done. So now suppose that $|W_{\mathbf{x}_1}^{B(y_0)}| > 1$ and let $(v^1, \cdots, v^r)$ be the subbasis of $\mathbf{x}_1$ consisting of those $v \in \mathbf{x}_1$ such that $\mathsf{P}^v$ and $\mathsf{P}^{y_0}$ anticommute. Then, by our recursive relation in Proposition 9, we have

$$U^\Lambda = \frac{1}{2^{r/2}} \left[ \prod_{j=1}^{r} \left( \mathsf{I}^{\otimes n} + (-1)^{\sigma_1(v^j)} i \mathsf{P}^{v^j} \right) \right] \cdot (-1)^\tau \mathsf{P}^{y_0}$$

Expanding this, we get the identity in Equation (4). □

We generalize this below in Proposition 12, showing that a depth-$d$ presentation expands into a uniform-magnitude linear combination of depth-$(d-1)$ presentations. We first prove the following intermediate result, which is a just a generalization of the algebraic relation in Proposition 10.

*Lemma 4:* Let $d \geq 0$ and $\Lambda, \Lambda'$ be two $n$-qubit depth-$d$ presentations differing only on their phased outer vectors, i.e.

$$\Lambda = ((\mathbf{x}_d, \sigma_d), \ldots, (\mathbf{x}_1, \sigma_1), (y_0, \tau))$$

$$\Lambda' = \big((\mathbf{x}_d, \sigma_d), \ldots, (\mathbf{x}_1, \sigma_1), (y_0', \tau')\big)$$

Finally, we define $\Theta$ to be the following $n$-qubit depth-$d$ presentation:

$$\Theta := \big((\mathbf{x}_d, \sigma_d), \ldots, (\mathbf{x}_1, \sigma_1), (y_0 \oplus y_0', \tau \oplus \tau')\big)$$

Then

$$U^\Lambda U^{\Lambda'} = i^{[y_0, y_0']} U^\Theta$$

In particular, $U^\Lambda$ and $U^{\Lambda'}$ commute if and only if $\mathsf{P}^{y_0}$ and $\mathsf{P}^{y_0'}$ commute, and otherwise they anticommute.

*Proof:* We prove this by induction on $d$, where the $d = 0$ case is just the relation in Proposition 10, so let $d \geq 1$ and suppose the induction hypothesis holds for $d-1$. As in Proposition 9, we define the following:

1) We let $\Lambda_*$ and $\Lambda_*'$ denote the following respective depth-$(d-1)$ presentations:

$$\Lambda_* := ((\mathbf{x}_d, \sigma_d), \ldots, (\mathbf{x}_2, \sigma_2), (y_0, \tau))$$

$$\Lambda_*' := \big((\mathbf{x}_d, \sigma_d), \ldots, (\mathbf{x}_2, \sigma_2), (y_0', \tau')\big)$$

2) For each basis vector $v$ of $\mathbf{x}_1$, we let $\Delta_v$ denote the depth-$(d-1)$ presentation $\Delta_v := ((\mathbf{x}_d, \sigma_d), \cdots, (\mathbf{x}_2 \sigma_2), (v, \sigma_1(v)))$.

Thus, we get the following expression for $U^\Lambda U^{\Lambda'}$.

$$\left( \prod_{\substack{v \in \mathbf{x}_1 \\ B(v, y_0) = 1}} \frac{\mathsf{I}^{\otimes n} + i U^{\Delta_v}}{\sqrt{2}} \right) U^{\Lambda_*} \left( \prod_{\substack{v \in \mathbf{x}_1 \\ B(v, y_0') = 1}} \frac{\mathsf{I}^{\otimes n} + i U^{\Delta_v}}{\sqrt{2}} \right) U^{\Lambda_*'}$$

Now, for each basis vector $v$ in $\mathbf{x}_1$ with $B(v, y_0') = 1$, if we have $B(v, y_0) = 1$ as well, then $U^{\Delta_v}$ anticommutes with $U^{\Lambda_*}$. Since each $U^{\Delta_v}$ is Hermitian, we get the following expression for $U^\Lambda U^{\Lambda'}$:

$$\left( \prod_{\substack{v \in \mathbf{x}_1 \\ B(v, y_0) = 1 \\ B(v, y_0') = 0}} \frac{\mathsf{I}^{\otimes n} + i U^{\Delta_v}}{\sqrt{2}} \right) U^{\Lambda_*} \left( \prod_{\substack{v \in \mathbf{x}_1 \\ B(v, y_0') = 1 \\ B(v, y_0) = 0}} \frac{\mathsf{I}^{\otimes n} + i U^{\Delta_v}}{\sqrt{2}} \right) U^{\Lambda_*'}$$

Now, by our induction hypothesis, each unitary of $(U^{\Delta_v} : B(v, y_0) = 0)$ commutes with $U^{\Lambda_*}$, as $U^{\Delta_v}$ and $U^{\Lambda_*}$ have respective outer vectors $v$ and $y_0$. So we can propagate $U^{\Lambda_*}$ to the right to get

$$U^\Lambda U^{\Lambda'} = \left( \prod_{\substack{v \in \mathbf{x}_1 \\ B(v, y_0 \oplus y_0') = 1}} \frac{\mathsf{I}^{\otimes n} + i U^{\Delta_v}}{\sqrt{2}} \right) U^{\Lambda_*} U^{\Lambda_*'}$$

Now let $\Theta_*$ denote the depth-$(d-2)$ presentation $\big((\mathbf{x}_d, \sigma_d), \ldots, (\mathbf{x}_2, \sigma_2), (y_0 \oplus y_0', \tau \oplus \tau')\big)$. Applying our induction hypothesis to $U^{\Lambda_*} U^{\Lambda_*'}$, we get

$$U^\Lambda U^{\Lambda'} = i^{[y_0, y_0']} \left( \prod_{\substack{v \in \mathbf{x}_1 \\ B(v, y_0 \oplus y_0') = 1}} \frac{\mathsf{I}^{\otimes n} + i U^{\Delta_v}}{\sqrt{2}} \right) U^{\Theta_*} \tag{5}$$

But now, by one more application of the recursive relation of Proposition 9, we get that the the RHS of (5) is precisely $i^{[y_0, y_0']} U^\Theta$, so we are done. □

With Lemma 4 in hand, we can generalize Proposition 11 to Proposition 12. Note that, in the statement below, $\frac{\omega(\mathbf{x}_1, z) - [z, y_0]}{2}$ is indeed a well-defined element of $\mathbb{F}_2$, as $\omega(\mathbf{x}_1, z)$ and $[z, y_0]$ always have the same parity for $z \in W_{\mathbf{x}_1}^{B_{\mathbf{x}_1}(y_0)}$.

*Proposition 12:* Let $\Lambda = ((\mathbf{x}_d, \sigma_d), \ldots, (\mathbf{x}_1, \sigma_1), (y_0, \tau))$ be a depth-$d$ presentation, for $d \geq 1$. For each $z \in W_{\mathbf{x}_1}^{B_{\mathbf{x}_1}(y_0)}$, we associate to $z$ the depth-$(d-1)$ presentation $\Lambda_z$ defined as follows:

$$\Lambda_z := ((\mathbf{x}_d, \sigma_d), \ldots, (\mathbf{x}_2, \sigma_2), (z \oplus y_0, f(z)))$$

where

$$f(z) := \tau \oplus \sigma_1(z) \oplus \theta_{\mathbf{x}_1}(z) \oplus \frac{\omega(\mathbf{x}_1, z) - [z, y_0]}{2}$$

Then we have

$$U^\Lambda = \left( \frac{1}{\sqrt{2}} \right)^{|B_{\mathbf{x}_1}(y_0)|} \sum_{z \in W_{\mathbf{x}_1}^{B(y_0)}} U^{\Lambda_z}$$

*Proof:* We define the following:

1) As in Proposition 9, we let $\Lambda_*$ be the depth-$(d-1)$ presentation $((\mathbf{x}_d, \sigma_d), \ldots, (\mathbf{x}_2, \sigma_2), (y_0, \tau))$, and, for each basis vector $v$ of $\mathbf{x}_1$, let $\Delta_v$ denote the depth-$(d-1)$ presentation $((\mathbf{x}_d, \sigma_d), \ldots, (\mathbf{x}_2, \sigma_2), (v, \sigma_1(v)))$.

2) For each $z \in W_{\mathbf{x}_1}^{B_{\mathbf{x}_1}(y_0)}$, we let $S(z)$ denote the set of basis vectors of $(v \in \mathbf{x}_1 : B(v, y_0) = 1)$ appearing with nonzero coefficient in the unique $\mathbb{F}_2$-linear combination of the $\mathbf{x}_1$-vectors summing to $z$.

Now, it suffices to prove that, for each $z \in W_{\mathbf{x}_1}^{B_{\mathbf{x}_1}(y_0)}$, the unitary $U^{\Lambda_z}$ decomposes into the following product where the individual factors corresponding to the elements of $S(z)$ pairwise commute (and so the product is well-defined):

$$U^{\Lambda_z} = i^{\omega(\mathbf{x}_1, v)} \left( \prod_{v \in S(z)} U^{\Delta_v} \right) U^{\Lambda_*} \tag{6}$$

It follows from Lemma 4 that Equation (6) holds, and, in particular, that the individual unitaries of $\{U^{\Delta_v} : v \in S(z)\}$ indeed pairwise commute, as we are only manipulating depth-$(d-1)$ presentations that differ from each other only by their phased outer vectors. With Equation (6) in hand, the desired result just follows by expanding the product in the recursive relation from Proposition 9. □

Proposition 12 shows that $U^{\Lambda}$ expands into a uniform linear combination of $2^{|B_{\mathbf{x}_1}(y_0)|}$ depth-$(d-1)$ presentations indexed (via their respective outer vectors) by elements of the affine space $W_{\mathbf{x}_1}^{B_{\mathbf{x}_1}(y_0)} \oplus y_0$. When we repeat this $d-1$ more times, we obtain a linear combination of depth-zero presentations, which are Paulis. For any $c \in \mathbb{F}_2^{2n}$, the contributions to the coefficient of $\mathsf{P}^c$ in $U^{\Lambda}$ are given by the sequences $(y_0, y_1, \ldots, y_d) \in \mathbb{F}_2^{2n(d+1)}$, where $y_d = c$ and, for each $j = 1, \ldots, d$, we have $y_j \in y_{j-1} \oplus W_{\mathbf{x}_j}^{B_{\mathbf{x}_j}(y_{j-1})}$. Each such string encodes a magnitude, namely:

$$\left( \frac{1}{\sqrt{2}} \right)^{\sum_{j=1}^{d} |B_{\mathbf{x}_j}(y_{j-1})|}$$

And it encodes a phase, namely:

$$\begin{aligned} \tau \oplus \bigoplus_{j=1}^{d} \left( \sigma_j(y_j \oplus y_{j-1}) \oplus \theta_{\mathbf{x}_j}(y_j \oplus y_{j-1}) \right) \\ \oplus \bigoplus_{j=1}^{d} \left( \frac{\omega(\mathbf{x}_j, y_j \oplus y_{j-1}) - [y_j \oplus y_{j-1}, y_{j-1}]}{2} \right) \end{aligned} \tag{7}$$

Then, defining the set $\mathrm{supp}(\Lambda)$ as:

$$\Big\{ \vec{y} = (y_1, \ldots, y_d) \in \mathbb{F}_2^{2nd} : y_j \in y_{j-1} \oplus W_{\mathbf{x}_j}^{B_{\mathbf{x}_j}(y_{j-1})} \text{ for } 1 \le j \le d \Big\}$$

and defining $\rho(\Lambda|\cdot) : \mathrm{supp}(\Lambda) \to \mathbb{F}_2$ to be the function assigning $(y_1, \ldots, y_d)$ to the expression in (7), we have the following expression for $U^{\Lambda}$:

$$\sum_{c \in \mathbb{F}_2^{2n}} \left( \sum_{\substack{\vec{y} \in \mathrm{supp}(\Lambda) \\ y_d = c}} (-1)^{\rho(\Lambda|\vec{y})} \left( \frac{1}{\sqrt{2}} \right)^{\sum_{j=1}^{d} |B_{\mathbf{x}_j}(y_{j-1})|} \right) \mathsf{P}^c \tag{8}$$

## X. Codes Generated by 1-Remainder Matrices

To continue, we first recall the following coding theory preliminaries.

*Definition 24:*

- For any integer $q$ of the form $p^m$, for $p$ prime, we let $\mathbb{F}_q$ denote the unique field of order $q$. A length $n$ $q$-ary code is a vector subspace $V \subseteq \mathbb{F}_q^n$, whose elements are called *codewords*. In particular, if $q = 2$, then $V$ is called a *binary code*. If $V$ has dimension $k$ over $\mathbb{F}_q$, then it is called an $[n, k]_q$-code. We call $n$ the *length* of $V$.
- For any $v \in \mathbb{F}_q^n$, the *Hamming weight* of $v$, denoted $|v|$, is the number of entries of $v$ that are nonzero, when $v$ is regarded as a length-$n$ string of characters from $\mathbb{F}_q$. In particular, this coincides with the definition in the $q = 2$ case introduced in Section IV. For $x, y \in V$, the *distance* between $x$ and $y$ is $|x - y|$. The *minimum distance* of $V$ is the minimum of the distances between any two distinct codewords of $V$.
- Given a $q$-ary matrix $G$ with $n$ columns, we say that $G$ *generates* $V$ if $V$ is the span of the row vectors of $G$. We call $G$ a *generator matrix* if its rows are linearly independent.
- We associate to $V$ a univariate polynomial $\mathrm{wt}_V(x) \in \mathbb{Z}[x]$, called the *weight enumerator polynomial* of $V$, defined as

$$\mathrm{wt}_V(x) := \sum_{j=0}^{n} a_j x^j$$

where, for each $0 \le j \le n$, we have $a_j := \#\{v \in V : |v| = j\}$. The coefficient sequence $(a_n, a_{n-1}, \ldots, a_0)$ is also called the *weight distribution* of $V$.

In some contexts, the word *code* refers to any subset of $\mathbb{F}_q^n$, but, in this paper, we are exclusively concerned with linear codes, i.e. the term code always refers to a vector subspace of $\mathbb{F}_q^n$. In general, computing the weight distribution of a binary code, or even just determining which $a_j$ are nonzero, is very difficult. A well-known hardness result from [22] is the following.

*Theorem 7:* The following problem is NP-complete: On input an integer $t \ge 0$ and a $k \times n$ binary matrix $A$, decide whether there is an $x \in \mathbb{F}_2^k$ such that $|x| = t$ and $xA = \mathbf{0}_n$.

As noted in [23], this also implies Theorem 8 below.

*Theorem 8:* The following decision problem, which we denote BINARY-WEIGHT, is NP-complete: On input an integer $t \ge 0$ and a binary matrix whose rows generate binary code $V$, decide whether there is an $x \in V$ with $|x| = t$.

Theorem 8 was later famously improved by Vardy ([24]), who showed the stronger statement that MINIMUM DISTANCE is NP-complete, i.e. the problem of deciding, on input an integer $t \ge 0$ and a binary matrix whose rows generate a binary code $V$ whether there is a nonzero $x \in V$ with $|x| \le t$, but Theorem 8 is enough for our purposes. We also recall the following definitions.

*Definition 25:* Let $V, V'$ be two length-$n$ codes over $\mathbb{F}_q$. An *isometry* between $V, V'$ is an $\mathbb{F}_q$-linear map $L : V \to V'$ which is distance-preserving, i.e. $|L(x) - L(y)| = |x - y|$ for any $x, y \in V$. We say that $V$ and $V'$ are *equivalent* if there is a a linear isometry from $V$ to $V'$. We say that $V$ and $V'$ are *permutationally equivalent* if $V'$ can be obtained from $V$ by some permutation of the components of $\mathbb{F}_q^n$.

All permutations of $\mathbb{F}_q^n$ are isometries $\mathbb{F}_q^n \to \mathbb{F}_q^n$, but the converse is not true, unless $q = 2$. Over $\mathbb{F}_2$, two length-$n$ codes are equivalent if and only if they are permutationally equivalent. Note that any two equivalent codes have the same weight distribution. However, the converse is not true, ever for $q = 2$. There are inequivalent binary codes with the same weight distribution. We now note that the weight enumerator polynomial obeys the following useful relationship.

*Observation 5:* For any $k \geq 1$ and binary code $V \subseteq \mathbb{F}_2^n$, letting $V^k$ be the binary code $\{(v, \ldots, v) : v \in V\} \subseteq \mathbb{F}_2^{nk}$, we have $\mathrm{wt}_{V^k}(x) = \sum_{v \in V} x^{k|v|} = \mathrm{wt}_V(x^k)$.

We also note the following:

*Theorem 9:* The following problem is #*P*-hard: On input a binary matrix generating binary code $V$, output the weight distribution of $V$.

*Proof:* A result of Jaeger, Vertigan, and Welsh ([25]) is that, for an arbitrary graphic matroid $M$, in the worst case, it is #P-hard to evaluate the bivariate Tutte polynomial $T(M; x, y)$ at every point in the $(x, y)$-plane except over the hyperplane $(x-1)(y-1) = 1$ and at four additional special points. Graphic matroids are a special case of binary matroids, and, for any binary matroid $M$, corresponding to $[n, k]_2$-code $V$, we have the following relationship from Green ([26])

$$\mathrm{wt}_V(x) = (x-1)^k T\left(M; \frac{x+1}{x-1}, x\right)$$

(Also see [27]. In particular, the problem of even approximating the evaluation of a weight enumerator of a binary code is #P-hard). □

To prove Theorem 6, given a binary code, we need to embed it in a depth-3 presentation $\Lambda$ in such a way that the phase function $\rho(\Lambda|\cdot)$ behaves nicely. In particular, we want to embed it in such a way that $\rho(\Lambda|\cdot)$ is identically zero. This is the motivation for Proposition 13, where we consider a special family of binary codes which permit such an embedding, where BINARY WEIGHT still remains NP-hard when we restrict ourselves to codes from this family.

*Definition 26:*

1) We define *1-remainder* matrix to be a binary matrix of the form $[I_k|I_k|\cdots|I_k|P|P|\cdots|P]$, where $P$ is an arbitrary $k$-row matrix, the number of $P$-blocks is 0 mod 4, and the number of $I_k$-blocks is 1 mod 4. Note that $[I_k|I_k|\cdots|I_k|P|P|\cdots|P]$ always has row-rank $k$.
2) Given an algebraic $\alpha \in \mathbb{C}$ as a parameter, we define $1\mathrm{MOD4}(\alpha)$ to be the following functional problem: Given as input a 1-remainder generator matrix $G$, generating binary code $V \subseteq \mathbb{F}_2^n$, output the value of $\mathrm{wt}_V(\alpha)$.
3) We also define a decision Problem 1MOD4-WEIGHT as follows: Given as input an integer $t \geq 0$ and 1-remainder matrix $G$ generating a binary code $V$, decide whether there is an $x \in V$ with $|x| = t$.

We then have the following:

*Proposition 13:*

1) For any algebraic $\alpha \in \mathbb{C}$ with $|\alpha| \neq 0, 1$, *1MOD4*$(\alpha)$ is #P-hard; AND
2) *1MOD4-WEIGHT* is NP-hard. In particular, there is a polynomial-time reduction from BINARY-WEIGHT to *1MOD4-WEIGHT*.

*Proof:* Let $P$ be a $k$-row generator matrix for an arbitrary binary code $V$ of length $n$. For each $m > 0$, let $H_m$ be the $k \times (k + 4mn)$ matrix $[I_k|P|\cdots|P]$ with $4m$ blocks of $P$, let $X_m \subseteq \mathbb{F}_2^{k+4mn}$ be the binary code generated by $H_m$, and let $A_m := (a_{m,k+4mn}, \ldots, a_{m,1}, a_{m,0})$ denote the coefficient sequence of $\mathrm{wt}_{X_m}(x)$.

Let $\mathbf{P}_\alpha$ be a procedure solving $1\mathrm{MOD4}(\alpha)$. We first claim that, for any $m > 0$, by calling $\mathbf{P}_\alpha$ $\mathrm{poly}(n, m)$ times on codes of length $\mathrm{poly}(n, m)$, we can recover all of $A_m$. For each $\ell > 0$, let $X_m^\ell := \{(v, \ldots, v) \in \mathbb{F}_2^{\ell(k+4mn)} : v \in X_m\}$, with each codeword repeated $\ell$ times. Note that, for $\ell \equiv 1 \bmod 4$, the code $X_m^\ell$ is equivalent to the binary code generated by the 1-remainder matrix $[I_k|\cdots|I_k|P|\cdots|P]$ with $\ell$ copies of $I_k$ and $4m\ell$ copies of $P$, so we can evaluate $\mathrm{wt}_{X_m^\ell}(\alpha) = \mathrm{wt}_{X_m}(\alpha^\ell)$. Since $|\alpha| \neq 0, 1$, we only need to call $\mathbf{P}_\alpha$ $O(k + 4mn)$ times on input binary codes of length $O((k+4mn)^2)$, in order to evaluate $\mathrm{wt}_{X_m}(x)$ at enough distinct points to efficiently recover $A_m$, as claimed.

Now, let $(b_n, b_{n-1}, \ldots, b_0)$ be the coefficient sequence of $\mathrm{wt}_V(x)$. Then:

a) to prove 1), it suffices to show that, for $m = O(\mathrm{poly}(n))$ sufficiently large, given access to $A_m$, we can efficiently recover $(b_n, b_{n-1}, \ldots, b_0)$.
b) Likewise, in order to prove 2), it suffices to show that, for $m = O(\mathrm{poly}(n))$ sufficiently large, given any $0 \leq i \leq n$, we can find a $0 \leq j \leq k + 4mn$ such that $b_i = 0$ if and only if $a_{m,j} = 0$.

We prove both of these together. For any integers $r, s$, we define auxiliary variables

$$x_{m,r,s} := \#\{(v, w) \in X_m : v \in \mathbb{F}_2^k,\ w \in \mathbb{F}_2^{4mn} \text{ s.t } |v| = r \text{ and } |w| = s\}$$

Thus, for each $0 \leq i \leq n$ and any $m > 0$, we have

$$b_i = \sum_{r=0}^{k} x_{m,r,4im}$$

For any $m > 0$, we have the following system of equations:

$$\mathcal{S}_m := \left\{ \sum_{\substack{0 \leq r \leq k \\ r \equiv \ell \bmod 4m}} x_{m,r,\ell-r} = a_{m,\ell} \text{ for all } 0 \leq \ell \leq k + 4mn \right\}$$

For $m > k/4$, we get the following: For any $\ell \geq 0$, we have $\#\{0 \leq r \leq k : r \equiv \ell \bmod 4m\} \leq 1$ (in particular, some of the equations in $\mathcal{S}_m$ have an empty LHS), and, for each $0 \leq i \leq n$, at least one term of the form $x_{m,r,4im}$ appears in $\mathcal{S}_m$. This gives us both a) and b). □

## XI. Relating Coefficients of $U^\Lambda$ to Weight Enumerator Polynomials

In this section, we finish the proof of Theorem 6. We temporarily abstract away from the setting of linear combinations of Paulis and instead look at formal sums of $2^n$ variables over $\mathbb{R}$, which are easier to study.

*Definition 27:* We define a $(d, n)$-*branching* to be a tuple $\mathcal{A} = ((\mathbf{x}_d, Q_d), \cdots, (\mathbf{x}_1, Q_1))$, where, for each $1 \leq j \leq d$,

1) $\mathbf{x}_j$ is a (possibly empty) ordered basis for a subspace of $\mathbb{F}_2^n$.

2) $Q_j$ is a linear map $\mathbb{F}_2^n \to \mathbb{F}_2^{|\mathbf{x}_j|}$ and furthermore, $W_{\mathbf{x}_j} \subseteq \ker(Q_j)$.

We let $\mathbb{R}[P_x : x \in \mathbb{F}_2^n]$ denote the $\mathbb{R}$-module of formal sums in $2^n$ variables, one for each element of $\mathbb{F}_2^n$. We now associate to each $(d,n)$-branching a function $\Phi_{\mathcal{A}} : \mathbb{F}_2^n \to \mathbb{R}[P_x : x \in \mathbb{F}_2^n]$ which assigns each $z \in \mathbb{F}_2^n$ an element of $\mathbb{R}[P_x : x \in \mathbb{F}_2^n]$. The function $\Phi_{\mathcal{A}}$ is defined recursively as follows:

1) If $\mathcal{A} = ()$ is a 0-branching, then $\Phi_{\mathcal{A}}(y) = P_y$.
2) Let $d \geq 1$ and suppose that, for any $0 \leq d' < d$ and $d'$-branching $\mathcal{A}'$, $\Phi_{\mathcal{A}'}$ is already defined. Let $\mathcal{A} = ((\mathbf{x}_d, Q_d), \dots, (\mathbf{x}_1, Q_1))$ be a $d$-branching and let $\mathcal{A}' := ((\mathbf{x}_d, Q_d), \dots, (\mathbf{x}_2, Q_2))$. We set

$$\Phi_{\mathcal{A}}(y) := \frac{1}{2^{|Q_1(y)|/2}} \left( \sum_{z \in W_{\mathbf{x}_1}^{Q_1(y)} \oplus y} \Phi_{\mathcal{A}'}(z) \right)$$

Given a $\Psi \in \mathbb{R}[P_u : u \in \mathbb{F}_2^n]$ and $q \in \mathbb{F}_2^n$, we let $\mathrm{co}(\Psi, q)$ denote the coefficient of $P_q$ in $\Psi$. By Equation (8), we immediately have the following:

*Observation 6:* Let $\Lambda = ((\mathbf{x}_d, \sigma_d), \dots, (\mathbf{x}_1, \sigma_1), (y_0, \tau))$ be a depth-$d$ presentation, and suppose that $\rho(\Lambda|\cdot)$ is identically zero. Let $\mathcal{A}$ be the $(2n, d)$-branching $((\mathbf{x}_d, B_{\mathbf{x}_d}), \dots, (\mathbf{x}_1, B_{\mathbf{x}_1}))$. Then, for each $z \in \mathbb{F}_2^{2n}$, we have $\langle U^\Lambda, \mathsf{P}^z \rangle = \mathrm{co}(\Phi_{\mathcal{A}}(y_0), z)$.

Now, we have the following. In particular, the coefficients of a $(2,n)$-branching can be calculated in polynomial time by Gaussian elimination.

*Proposition 14:* Given a $(2,n)$-branching $\mathcal{A} = ((\mathbf{x}_2, Q_2), (\mathbf{x}_1, Q_1))$, we have, for any $q, y \in \mathbb{F}_2^n$,

$$\mathrm{co}(\Phi_{\mathcal{A}}(y), q) = \frac{|(y \oplus W_{\mathbf{x}_1}^{Q_1(y)}) \cap (q \oplus W_{\mathbf{x}_2}^{Q_2(q)})|}{2^{|Q_1(y)|/2} \cdot 2^{|Q_2(q)|/2}}$$

*Proof:* We have

$$\Phi_{\mathcal{A}}(y) = \frac{1}{2^{|Q_1(y)|/2}} \cdot \left( \sum_{z \in W_{\mathbf{x}_1}^{Q_1(y)} \oplus y} \frac{1}{2^{|Q_2(z)|/2}} \cdot \left( \sum_{x \in W_{\mathbf{x}_2}^{Q_2(z)} \oplus z} P_x \right) \right)$$

Let $M := W_{\mathbf{x}_1}^{Q_1(y)} \oplus y$ and let $X := \{z \in M : q \in W_{\mathbf{x}_2}^{Q_2(z)} \oplus z\}$. We show now that $X = M \cap (q \oplus W_{\mathbf{x}_2}^{Q_2(q)})$. Let $z \in X$. Thus, there exist $u \in W_{\mathbf{x}_2}^{Q_2(z)}$ such that $q = z \oplus u$. Since $u \in \ker(Q_2)$, we get $Q_2(z) = Q_2(q)$. Let $V := W_{\mathbf{x}_2}^{Q_2(q)}$. In particular, we get $X = \{z \in M : \exists u \in V \text{s.t } q = z \oplus u\}$. Let $f \in X$. By definition, $f \in M$, and there exists a $u \in V$ such that $q = f \oplus u$. Thus, $f \in q \oplus V$, and $X \subseteq (q \oplus V) \cap M$. Conversely, if $f \in M$ with $f \oplus q \in V$, then, by definition, $f \in X$. So we indeed have $X = M \cap (q \oplus W_{\mathbf{x}_2}^{Q_2(q)})$. It follows that, for each $z \in X$, we have $Q_2(z) = Q_2(q)$, and we get

$$\mathrm{co}(\Phi_{\mathcal{A}}(y), q) = \frac{1}{2^{|Q_1(y)|/2}} \left( \sum_{z \in X} \frac{1}{2^{|Q_2(z)|/2}} \right) = \frac{|X|}{2^{|Q_1(y)|/2} \cdot 2^{|Q_2(q)|/2}}$$

as desired. □

We now move on to $(3,n)$-branchings:

*Proposition 15:* Let $\mathcal{A} = ((\mathbf{x}_3, Q_3), (\mathbf{x}_2, Q_2), (\mathbf{x}_1, Q_1))$ be a $(3,n)$-branching. Then, for any $y, q \in \mathbb{F}_2^{2n}$, letting $M := W_{\mathbf{x}_1}^{Q_1(y)} \oplus y$ and $N := W_{\mathbf{x}_3}^{Q_3(q)} \oplus q$, we have

$$\mathrm{co}(\Phi_{\mathcal{A}}(y), q) = \frac{1}{\sqrt{|M| \cdot |N|}} \left( \sum_{z \in M} \frac{|(z \oplus W_{\mathbf{x}_2}^{Q_2(z)}) \cap N|}{2^{|Q_2(z)|/2}} \right)$$

In particular, if both $W_{\mathbf{x}_2} \cap W_{\mathbf{x}_3} = \{\mathbf{0}\}$ and $M \subseteq N$, then

$$\Phi_{\mathcal{A}}(y), q) = \frac{1}{\sqrt{|M| \cdot |N|}} \sum_{z \in M} \frac{1}{2^{|Q_2(z)|/2}}$$

*Proof:* Let $\mathcal{A}' = ((\mathbf{x}_3, Q_3), (\mathbf{x}_2, Q_2))$. We then have

$$\Phi_{\mathcal{A}}(y) = \frac{1}{2^{|Q_1(y)|/2}} \cdot \left( \sum_{z \in M} \Phi_{\mathcal{A}'}(z) \right)$$

Thus, by Proposition 14, we have

$$\mathrm{co}(\Phi_{\mathcal{A}}(y), q) = \frac{1}{2^{|Q_1(y)|/2}} \cdot \left( \sum_{z \in M} \mathrm{co}(\Phi_{\mathcal{A}'}(z), q) \right) = \frac{1}{2^{|Q_1(y)|/2} \cdot 2^{|Q_3(q)|/2}} \left( \sum_{z \in M} \frac{|(z \oplus W_{\mathbf{x}_2}^{Q_2(z)}) \cap N|}{2^{|Q_2(z)|/2}} \right)$$

as desired. Now suppose that both $W_{\mathbf{x}_2} \cap W_{\mathbf{x}_3} = \{\mathbf{0}\}$ and $M \subseteq N$. We just need to check that, for each $z \in M$, we have $|(z \oplus W_{\mathbf{x}_2}^{Q_2(z)}) \cap N| = 1$. We have $|(z \oplus W_{\mathbf{x}_2}^{Q_2(z)}) \cap N| \geq 1$, since $z$ lies in both sets. Conversely, if there exist $u, u' \in N$ with $u, u' \in z \oplus W_{\mathbf{x}_2}^{Q_2(z)}$, then $u \oplus u' \in W_{\mathbf{x}_3} \cap W_{\mathbf{x}_2}$, so $u = u'$ and $|(z \oplus W_{\mathbf{x}_2}^{Q_2(z)}) \cap N| \leq 1$. □

We need two more reductions to complete the proof of Theorem 6. We first need the following simple observation.

*Lemma 5:* Let $\mathbf{x}$ be an ordered basis for either an $\mathsf{X}$-subspace or a $\mathsf{Z}$-subspace of $\mathbb{F}_2^{2n}$. For any $y \in \mathrm{Anti}(\mathbf{x})$, we define the following map $f(y, \cdot) : W_{\mathbf{x}} \to \mathbb{F}_2$

$$f(y, \cdot) : u \to \frac{\omega(\mathbf{x}, u) - [u, y]}{2}$$

If $\mathbf{x}$ consists of standard basis vectors then $f(y, \cdot)$ is identically zero over $W_{\mathbf{x}}$ for every $y \in \mathrm{Anti}(\mathbf{x})$.

*Proof:* Say without loss of generality that $W_{\mathbf{x}}$ is a $\mathsf{Z}$-subspace and, in particular, $\mathbf{x} = \mathbf{E}_{\mathsf{Z}}^r$ for some $1 \leq r \leq n$. Thus, for any $u \in W_{\mathbf{x}}$ and $y \in \mathbb{F}_2^{2n}$, we have $[u, y] = u_{\mathsf{Z}} \cdot y_{\mathsf{X}}$ as binary strings whose dot products are read as integers. Let $S := \{1 \leq j \leq r : (y_{\mathsf{X}})_j = 1\}$. Since $y \in \mathrm{Anti}(\mathbf{x})$, each $S \cap \{j\}$ is odd, so $S = [r]$. Thus, for any $T \subseteq [r]$, we have

$$f\left(y, \bigoplus_{j \in T} e_{\mathsf{Z}}^j\right) = \frac{|T| - |S \cap T|}{2} = 0$$

as desired. □

*Proposition 16:* Given as input a 1-remainder matrix $G$ generating binary code $V \subseteq \mathbb{F}_2^n$, we can efficiently construct a $C \in \mathcal{T}_n^3$, with $|C| = O(\mathrm{poly}(n))$ such that

$$\left\langle C\mathsf{Z}_1 C^\dagger, \mathsf{Z}_1 \right\rangle = \frac{1}{2^{n/2}\sqrt{|V|}} \mathrm{wt}_V\left(\frac{1}{\sqrt{2}}\right)$$

*Proof:* Let $G = [I_k|\cdots|I_k|P|\cdots|P]$ for some arbitrary matrix $P$, where $G$ has $r$ blocks of $I_k$ and $s$ blocks of $P$. We define a mapping $V \to \mathbb{F}_2^{2n}$ by $v \to (v, \mathbf{0}_n)$. Let $\mathbf{u}$ be an ordered basis, obtained from the rows of $G$, for the corresponding $\mathsf{Z}$-subspace of $\mathbb{F}_2^{2n}$. Let $y_0 := (\mathbf{0}_n|\mathbf{1}_n)$, i.e. $\mathsf{P}^{y_0} = \mathsf{X}^{\otimes n}$. Now, $W_{\mathbf{u}}$ is an isotropic subspace of $\mathbb{F}_2^{2n}$, and we set $\Lambda := ((\mathbf{E}^n_{\mathsf{Z}}, \mathbf{0}), (\mathbf{E}^n_{\mathsf{X}}, \mathbf{0}), (\mathbf{u}, \mathbf{0}), (y_0, 0))$. Recall that each of $\theta_{\mathbf{u}}$ and $\theta_{\mathbf{E}^n_{\mathsf{X}}}$ and $\theta_{\mathbf{E}^n_{\mathsf{Z}}}$ is identically zero by Observation 4. Furthermore, $y_0 \in \mathrm{Anti}(\mathbf{E}^n_{\mathsf{Z}})$, and it follows from Lemma 5 that, for $(y_1, y_2, y_3) \in \mathrm{supp}(\Lambda)$, we have

$$\rho(\Lambda|y_1, y_2, y_3) = \frac{\omega(\mathbf{u}, y_1 \oplus y_0) - [y_1 \oplus y_0, y_0]}{2}$$

Since $r \equiv 1 \bmod 4$ and $s \equiv 0 \bmod 4$, it follows that, for any $v \in W_{\mathbf{u}}$, we have $\omega(\mathbf{u}, v) \equiv |v| \bmod 4$ and $[v, y_0] = |v|$. Note that this also implies that $y_0 \in \mathrm{Anti}(\mathbf{u})$. Letting $M := W_{\mathbf{u}} \oplus y_0$, we get $M \subseteq W_{\mathbf{E}^n_{\mathsf{Z}}} \oplus y_0$, since $W_{\mathbf{u}} \subseteq W_{\mathbf{E}^n_{\mathsf{Z}}}$. It follows that $\rho(\Lambda|\cdot)$ is identically zero on $\mathrm{supp}(\Lambda)$ and, applying Proposition 15, where $N := W_{\mathbf{E}^n_{\mathsf{Z}}} \oplus y_0$, together with Observation 6, we have

$$\begin{aligned}\left\langle U^\Lambda, \mathsf{P}^{y_0}\right\rangle &= \tfrac{1}{\sqrt{|M|\cdot|N|}}\left(\textstyle\sum_{z \in M}\left(\tfrac{1}{\sqrt{2}}\right)^{|B_{\mathbf{E}^n_{\mathsf{X}}}(z)|}\right)\\ &= \tfrac{1}{2^{n/2}\sqrt{|V|}}\textstyle\sum_{v \in V}\left(\tfrac{1}{\sqrt{2}}\right)^{|v|}\end{aligned}$$

Applying Proposition 8 to express $U^\Lambda$ as a conjugate of $\mathsf{Z}_1$, and then just applying an appropriate Clifford shift (to shift from the $\mathsf{X}^{\otimes n}$-coefficient to the $\mathsf{Z}_1$-coefficient) as in Proposition 3, we are done. □

The reduction, above, together with Proposition 13, immediately gives us M1) of Theorem 6. For M2)-M3), we have the following.

*Proposition 17:* Given as input an integer $t \geq 0$ and a 1-remainder matrix $G$ generating binary code $V \subseteq \mathbb{F}_2^n$, we can efficiently construct an $O(\mathrm{poly}(n))$-qubit Clifford+T-circuit $F$ of T-depth $O(\log(n))$ and size $O\left(\mathrm{poly}(n)\right)$ such that $\langle F\mathsf{Z}_1F^\dagger, \mathsf{Z}_1\rangle = 0$ if and only if $\#\{v \in V : |v| = t\} = 0$.

*Proof:* Proposition 17 is trivial for $t > n$, so we fix a $0 \leq t \leq n$. We fix $\alpha := 1/\sqrt{2}$ and, for each integer $r \geq 0$, we let $\gamma_r := \mathrm{wt}_V(\alpha^r)$. Let $(b_n, b_{n-1}, \ldots, b_0)$ denote the weight distribution of $V$. We consider the $(n+1) \times (n+1)$-matrix

$$M := \begin{pmatrix} 1 & \alpha & \alpha^2 & \cdots & \alpha^n \\ 1 & \alpha^5 & \alpha^{10} & \cdots & \alpha^{5n} \\ 1 & \alpha^9 & \alpha^{18} & \cdots & \alpha^{9n} \\ \vdots & \vdots & \vdots & \ddots & \vdots \\ 1 & \alpha^{4n+1} & \alpha^{2(4n+1)} & \cdots & \alpha^{n(4n+1)} \end{pmatrix}$$

So we have the linear system

$$M\begin{pmatrix} b_0 \\ b_1 \\ b_2 \\ \vdots \\ b_n \end{pmatrix} = \begin{pmatrix} \gamma_1 \\ \gamma_5 \\ \gamma_9 \\ \vdots \\ \gamma_{4n+1} \end{pmatrix}$$

Now, $M$ is an invertible Vandermonde matrix, since $|\alpha| \neq 0, 1$. Let $(d_{ij})_{0 \leq i,j \leq n}$ denote the inverse. The entries of this matrix can just be computed by efficient matrix inversion, and each $d_{ij}$ is a ratio of two terms of $\mathbb{Z}[\sqrt{2}]$, each of which admits a length-$O(n^2)$ generalized base-$\sqrt{2}$ expression. However, this fraction does not necessarily admit any finite expansion in base-$\sqrt{2}$, so we clear denominators. Recall that, by the usual Vandermonde inversion formula, we have

$$d_{ij} = \frac{(-1)^{n-i}e_{n-i}\left(\{\alpha, \alpha^5, \ldots, \alpha^{4n+1}\} \setminus \{\alpha^{4j+1}\}\right)}{\prod_{\substack{0 \leq m \leq n \\ m \neq j}}(\alpha^{4j+1} - \alpha^{4m+1})}$$

where $e_m(y_1, \cdots, y_k)$ denotes the $m$th elementary symmetric function in variables $y_1, \ldots, y_k$, for $0 \leq m \leq k$. Thus, we define the term

$$\beta := \prod_{j=1}^{n}(1 - \alpha^{4j}) \neq 0$$

Then each $d_{tj}\beta$ is expressible in the form $d/2^\ell$, for $d \in \mathbb{Z}[\sqrt{2}]$ and $\ell \in \mathbb{Z}$, where $\ell = O(n^2)$ and $d$ admits a length-$O(n^2)$ generalized base-$\sqrt{2}$-expression. Furthermore, we have

$$\beta \cdot b_t = \beta d_{t0}\gamma_1 + \beta d_{t1}\gamma_5 + \beta d_{t2}\gamma_9 + \cdots + \beta d_{tn}\gamma_{4n+1}$$

We now show that, for any $0 \leq j \leq n$, we can efficiently construct both an $\ell_j \in \mathbb{Z}$, with $\ell_j = O(n^2)$, and an $O(\mathrm{poly}(n))$-qubit circuit $C_j$ of T-depth $O(\log(n))$ and size $|C_j| = O(\mathrm{poly}(n))$, where $\langle C_j\mathsf{Z}_1C^\dagger, \mathsf{Z}_1\rangle = \beta d_{tj}\gamma_{4j+1}/2^{\ell_j}$.

We first note that $\gamma_{4j+1}$ is precisely $\mathrm{wt}_{V'}(\alpha)$, where $V'$ is the length-$(4j+1)n$ code generated by $[G|\cdots|G]$ with $4j+1$ blocks of $G$, which is a 1-remainder matrix up to column permutation. Thus, combining Proposition 16 with Lemmas 1 and 3, we can indeed construct such a $C_j$. Now we just pad $C_0, \cdots, C_n$ with tensor products of identity until they have the same number $n' = O(n^2)$ of registers. We can also, again applying the arithmetic operations of Lemmas 1 and 3, transform each of the resulting circuits to scale their $\mathsf{Z}_1$-coefficients by powers of two so that they all have the same power of two in the denominator. Then we are done by one final application of Lemma 2. That is, we can construct an $F$ satisfying the requirements of Proposition 17, where, in particular, $\langle F\mathsf{Z}_1F^\dagger, \mathsf{Z}_1\rangle$ is proportional to $\beta \cdot b_t$ by some power of $1/\sqrt{2}$. □

Chaining the reductions in Propositions 17, 13, 6, and 5, we get M2)-M3) of Theorem 6. Note that M3) follows from M2), together with b of Proposition 6, and, for M2), we have the chain of reductions in Figure 4.

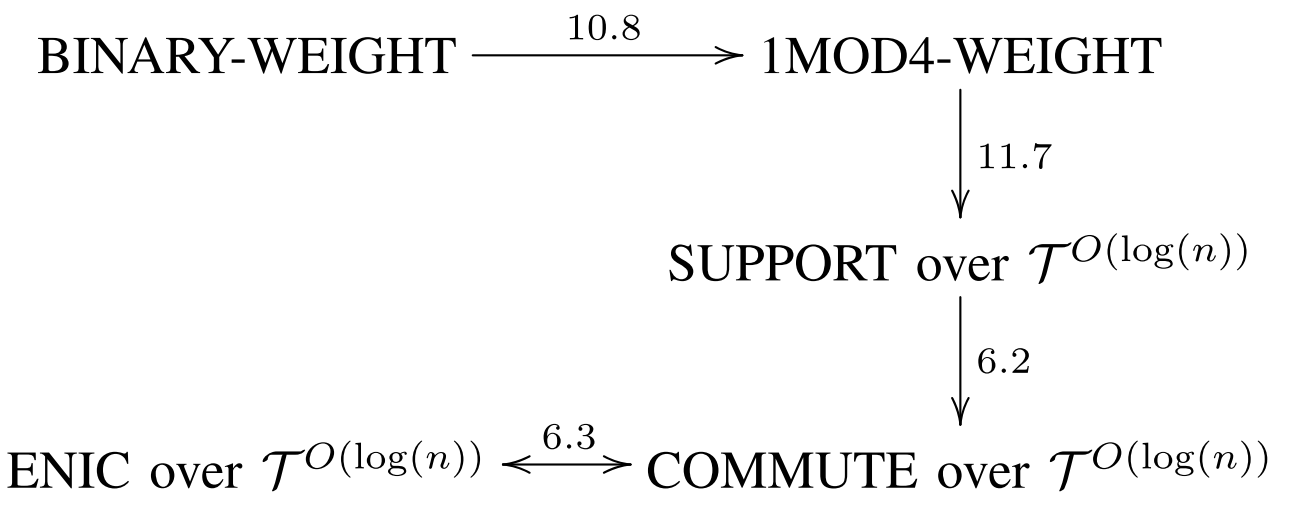

Fig. 4. Reduction from BINARY-WEIGHT to ENIC and COMMUTE over $\mathcal{T}^{O(\log(n))}$, where an arrow pointing from $A$ to $B$ denotes a polynomial-time reduction from $A$ to $B$, marked with the corresponding proposition.

## XII. Concluding Remarks and Open Questions

We have shown that ENIC, as well as the specified Pauli-expansion problems in Theorem 6, are NP-hard for circuits of logarithmic T-depth. On the other hand, we have also shown that ENIC can be decided in polynomial time over $\mathcal{T}^2$. This leaves a gap, so it is natural to ask whether ENIC remains hard at constant T-depth, or, alternatively, as in 2) of Question 1, if we can narrow the gap from below instead.

*Question 1:*

1) Does ENIC remain NP-hard when restricted to $\mathcal{T}^{O(1)}$?
2) Can we still find a $d > 2$ such that ENIC is polynomial-time over $\mathcal{T}^d$? What about $d = 3$? Given such a $d$, can the protocol of [1] be extended to $\mathcal{T}^d$, i.e. does this family admit efficient conjugate-encoding?

We also leave open the following natural question.

*Question 2:* Does Theorem 6 admit a qudit analogue for $d$ any prime power? In particular, given a $d$-ary code $V$ (under some suitable restrictions analogous to Definition 26), can we construct a qudit circuit $C$ and an element $P$ of the Heisenberg-Weyl group such that there is a coefficient of the expansion of $CPC^\dagger$ in the Heisenberg-Weyl basis which can be cast as the evaluation of $\mathrm{wt}_V(x)$ at some nonzero point of $\mathbb{C}$ outside the unit circle?

As Question 1 appears difficult, we now discuss a variant that seems easier. Recall that the proof of the NP-hardness of ENIC over $\mathcal{T}^{O(\log(n))}$ consists of two ingredients:

i) For any $d$, ENIC over domain $\mathcal{T}^{2d+O(\log(n))}$ is at least as hard as SUPPORT over domain $\mathcal{T}^d$.
ii) SUPPORT over domain $\mathcal{T}^{O(\log(n))}$ is NP-hard.

The proof of i) requires the implemention of a $(2n+1)$-Toffoli which in turn requires a T-depth of $\Theta(\log(n))$, but we suspect that ii) can be brought down from $\mathcal{T}^{O(\log(n))}$ to $\mathcal{T}^{O(1)}$:

*Question 3:* Is SUPPORT over $\mathcal{T}^{O(1)}$ still NP-*hard*? Is it at least as hard as binary code equivalence?

We suspect that the answer to the first (and thus also the second) part of Question 3 is yes. In particular, using our reductions, it would suffice to prove the NP-hardness of *formal code equivalence*, the problem of deciding whether two given binary codes have the same weight enumerator polynomial. More precisely, using Lemma 1 and Proposition 16, we get the following:

*Proposition 18:* Given as input a pair of 1-remainder matrices $G, G'$ generating respective binary codes $V, V' \subseteq \mathbb{F}_2^n$ of the same dimension $k$, we can efficiently construct a $(2n+1)$-qubit circuit Clifford+T-circuit of T-depth $\leq 5$, with $|C| = O(\mathrm{poly}(n))$, such that

$$\left\langle C\mathsf{Z}_1C^\dagger, \mathsf{Z}_1\right\rangle = \frac{1}{2^{(n+k+2)/2}}\left(\mathrm{wt}_V\left(\frac{1}{\sqrt{2}}\right) - \mathrm{wt}_{V'}\left(\frac{1}{\sqrt{2}}\right)\right)$$

Proposition 18, together with the argument of Proposition 13, is enough to show that any Turing machine that can efficiently decide SUPPORT over $\mathcal{T}^5$ can also efficiently decide formal code equivalence, as any two length-$n$ binary codes $V, V'$ have the same weight enumerator polynomial if and only if there are $n+1$ points $\ell \in \mathbb{N}$ such that $\mathrm{wt}_V\left(2^{-\ell/2}\right) = \mathrm{wt}_{V'}\left(2^{-\ell/2}\right)$. An affirmative answer to Question 3 would not by itself be enough to strengthen the result that ENIC is NP-hard over $\mathcal{T}^{O(\log(n))}$, but it would simplify the proof: We would no longer need to consider linear combinations of $O(n)$ terms of the form given in Proposition 16, which means we could simply dispense with all of Section VII and Proposition 17.

The relationship between SUPPORT and formal code equivalence discussed above is in turn the motivation for the weaker second part of Question 3, because the problems of code equivalence and formal code equivalance are closely related. The problem of deciding whether two binary codes are equivalent is well-known (see [28], [29], and [30]). As shown in [29], that problem is at least as hard as GRAPH ISOMORPHISM, but also unlikely to be NP-hard, unless the polynomial hierarchy collapses. Also, binary code equivalence reduces to the lattice isomorphism problem ([31]), which lies in SZK ([32]). So binary code equivalence is unlikely to be too hard, but also unlikely to be easy. The relationship between the problems of code equivalence and formal code equivalence is discussed by Sendrier in [33]. There are some heuristics from [33] which suggest that it is very likely that formal binary code equivalence is at least as hard as binary code equivalence (see Sections 4.1 and 5.2 of [33] and Section 3.1.1 of [28] for some discussion of these heuristics). However, to our knowledge, it remains open to produce a formal proof of this.

## Acknowledgment

The author would like to thank Andrej Bogdanov, Anne Broadbent, and Daniel Lovsted for interesting and productive discussion. He would also like to thank the anonymous reviewers for detailed and helpful feedback, which improved the clarity and organization of the article.

**Joshua Nevin** received the B.Sc. degree in mathematics from California Institute of Technology in 2013, the M.Sc. degree in mathematics from The University of British Columbia in 2015, and the Ph.D. degree in combinatorics and optimization from the Faculty of Mathematics, University of Waterloo, in 2021, under the supervision of Prof. Bruce Richter. He was subsequently a Post-Doctoral Fellow at the Faculty of Mathematics, University of Waterloo, again under the supervision of Bruce Richter. Following this, he was a Post-Doctoral Fellow at the Department of Mathematics, University of Ottawa, under the supervision of Prof. Anne Broadbent as part of the internal Quantum Security via Algebras and Representation Theory (QUASAR) Research Group and Ottawa node of the joint European-Canadian Foundations of Quantum Computational Advantage (FoQaCiA) Research Project.